\documentclass{emulateapj}
\usepackage{apjfonts}

\newcommand{\beq}{\begin{equation}}
\newcommand{\beqa}{\begin{eqnarray}}
\newcommand{\eeq}{\end{equation}}
\newcommand{\eeqa}{\end{eqnarray}}
\newcommand{\siml}{\lesssim}
\newcommand{\simg}{\gtrsim}

\shorttitle{Very High Lorentz Factor GRB Fireballs}
\shortauthors{Ioka}

\begin{document}

\title{
Very High Lorentz Factor Fireballs and Gamma-Ray Burst Spectra
}



\author{Kunihito Ioka
}

\affil{
KEK Theory Center
and the Graduate University for Advanced Studies (Sokendai), 
Tsukuba 305-0801, Japan
}



\begin{abstract}
Collisionless entrainment of the surrounding matter imports the relativistic 
baryon component in the Gamma-Ray Burst (GRB) fireball frame. 
We show that half the fireball energy can
be transferred from radiation to the comoving hot motions of baryons
under the photosphere.
The yet baryon-poor fireball can 
reexpand to a very high Lorentz factor 
(VHLF) $\Gamma \sim 10^3$--$10^6$ by its own relativistic collisionless 
pressure beyond the photosphere 
(so-called collisionless bulk acceleration), leading to internal and 
external shocks.
A simple synchrotron emission from the VHLF internal shocks produces
(i) the extra power-law spectral component with variability observed 
in the Fermi GeV bursts, up to the TeV range for the future Cherenkov 
Telescope Array (CTA), (ii) the GeV onset delay with a weak luminosity 
dependence $t_{\rm delay} \sim L^{-1/5}$, and (iii) the spectral break of 
GRB 090926 by the synchrotron cooling break or the maximum synchrotron 
cutoff limited by the dynamical time, not by the $e^{\pm}$ creation cutoff. 
The relativistic baryon component could also heat the photospheric thermal 
photons into the main GRB Band spectrum via $pp$, $p\gamma$ 
(Bethe-Heitler and photomeson), and Coulomb thermalization processes.
In this hot photosphere--internal--external shock model, we can predict the 
anticorrelation of $\sim$TeV neutrinos and GeV $\gamma$-rays, which may be 
detectable using IceCube. 
The spectral peak and luminosity (Yonetoku) relation is also reproduced
if the progenitor stars are nearly identical.
We also discuss the steep/shallow decay of early X-ray afterglows 
and short GRBs.
\end{abstract}





\section{Introduction}\label{sec:intro}

The cosmological Gamma-Ray Bursts (GRBs) are 
the most luminous objects in the Universe.
Although the physical processes at the
central engine are far from understood,
the fireball model is generally accepted
as the paradigm for producing the relativistic outflows
and high-energy emission
\citep{Cavallo:1978,p86,g86,sp90,mr93,Meszaros:1993ju,
Grimsrud:1998me,Meszaros:1999gb}.

However, the actual emission mechanism of prompt GRBs is still debated,
lacking a consistent picture.
The main problem is the high efficiency ($\simg 50\%$) 
of the GRB prompt emission,
defined by the GRB energy divided by the total energy including
the afterglow energy \citep{Zhang:2006uj,Ioka:2005zj}.
With the use of the internal shock model 
(the leading model for the prompt emission),
it is difficult to achieve a high efficiency
without a large dispersion in the Lorentz factor of the outflows
\citep{Kobayashi:1997jk,Kobayashi:2001iq,Beloborodov:2000nn}.
Even if it is achieved, a large dispersion in the Lorentz factor
makes it difficult to realize the observed spectral relations
\citep{Zhang:2002jt},
such as the Amati \citep{Amati:2009ts,Amati:2010rd} 
and Yonetoku relations \citep{Yonetoku:2003gi,Kodama:2008dq,Nava:2010rb},
\beqa
\varepsilon_{\rm peak} \simeq 600 
\left(\frac{L}{10^{53}\ {\rm erg}\ {\rm s}^{-1}}\right)^{1/2} {\rm keV},
\label{eq:yonetoku}
\eeqa
where $\varepsilon_{\rm peak}$ is the peak energy
of the observed broken power-law spectrum
(so-called the Band spectrum)
and $L$ is the apparent isotropic luminosity of the prompt emission
observed within an angle $<1/\Gamma$ of a jet axis.
Such a correlation is also satisfied 
within individual pulses \citep{Ohno:2008xk,Ghirlanda:2009pj}.
Another problem of the internal shock synchrotron model
is that the low-energy spectral slope becomes steeper than 
that observed,
owing to the intrinsic synchrotron spectrum and, even worse, 
owing to the fast electron cooling
\citep{Ghisellini:1999wu,Meszaros:1999gb}.

These difficulties of the internal shock models lead to
the reexamination of the original fireball model \citep{p86,g86},
in which photons are released as photospheric emission
when the fireball becomes optically thin
\citep{Thompson:1994,Rees:2004gt,Thompson:2006fp,
Meszaros:1999gb,dkk99,Meszaros:2002vh,Pe'er:2003ft,
Thompson:2005gt,Pe'er:2005kz,Ryde:2005pm,Giannios:2006jb,
Pe'er:2007xm,Giannios:2007yj,
Ioka:2007qk,Ghisellini:2007tm,Ryde:2008ir,
Beloborodov:2009be,Lazzati:2010af}.
The original problem is alleviated
by introducing the dissipation under the photosphere 
\citep{Rees:2004gt,Thompson:2006fp},
which can bring the thermal peak into the observed range 
in Eq.~(\ref{eq:yonetoku}).
The photosphere model can naturally achieve
the high efficiency and the hard low-energy spectrum.
The only flaw is that the spectrum tends to be thermal
without nonthermal tails observed in GRBs,
although a substantial fraction ($\sim 30\%$) 
of long GRBs may have thermal peaks \citep{Ryde:2005pm,Ryde:2008ir}.
The nonthermal tails could arise from
Comptonization of the thermal photons by electrons and positrons ($e^{\pm}$),
heated at dissipation, such as 
magnetic reconnection \citep{Giannios:2006jb,Giannios:2007yj}, 
neutron collisions \citep{dkk99,Beloborodov:2009be}, 
or repeated shocks \citep{Ioka:2007qk,Lazzati:2010af}.
However, it seems questionable that 
a subdominant or a different component rather than the thermal component
can supply the nonthermal energy
that is accidentally comparable to
the dominant thermal energy.
Nevertheless, the photosphere model has an advantage
that the peak energy $\varepsilon_{\rm peak}$
is stabilized as it is fixed by the temperature of the photosphere,
regardless of the dissipation mechanism.
The spectral relations in Eq.~(\ref{eq:yonetoku})
are more easily reproduced \citep{Rees:2004gt,Thompson:2006fp}
than the other attempts to solve the emission mechanism, such as
the jitter radiation \citep{Medvedev:1999tu,Medvedev:2000gu},
Klein-Nishina effect \citep{Derishev:2000dr,Bosnjak:2008bd,Wang:2009rp},
synchrotron self-Compton (SSC) \citep{Panaitescu:2000zv,Stern:2004xh},
bulk Compton \citep{Lazzati:1999cs,Ghisellini:2000qs,Lazzati:2003af},
decaying magnetic field \citep{Rossi:2002bf,Pe'er:2006yx},
and quasi-thermal Comptonization \citep{Ghisellini:1998jy,Asano:2009gi}.

Recently, the Fermi satellite, launched on 11 June 2008 
with the GBM (8 keV -- 40 MeV) and LAT ($\sim 20$ MeV -- 300 GeV)
detectors, has been used to observe $\sim$GeV $\gamma$-rays from GRBs,
providing interesting clues to the emission mechanism
\citep{Abdo:2009pg,Abdo:2009a,Abdo:2009b,
Abdo:2009nj,Abdo:2010gr,Abdo:2010a,LAT:2010us}.
The GeV events are increasing more than sixfold from the era of EGRET
that detected an 18 GeV photon 90 min after the 
burst in GRB 940217 \citep{Hurley:1994} and
a rising late GeV spectral component in GRB 941017 \citep{Gonzalez:2003}.
The main features of the Fermi bursts are summarized as follows:
\begin{itemize}
\item[(1)] In some Fermi/LAT bursts,
the Lorentz factor of the outflows is constrained in the 
relatively high range $\Gamma \simg 10^3$
so that the high-energy photons 
can avoid the annihilation by $e^{\pm}$ pair creation
\citep{Abdo:2009a,Abdo:2009pg,LAT:2010us}.
\item[(2)] Fermi found an additional spectral component at $\simg 10$ MeV
with comparable energy to that of the main Band component, 
at least, in short GRB 090510 \citep{LAT:2010us} 
and long GRB 090902B \citep{Abdo:2009pg,Ryde:2009wn}.
This extra component is fitted by a single power-law that
slightly rises in $\nu F_{\nu} \propto \nu^{0.1}$--$\nu^{0.4}$
and often extends below $\siml 20$ keV over $\simg 7$ energy digits.
(However, note that no other experiments have confirmed the
low-energy extension of the power-law component.)
\item[(3)] The high-energy ($>100$ MeV) emission lasts longer than 
the MeV emission in most LAT GRBs
\citep{Abdo:2009nj,Abdo:2009a,Abdo:2010gr,Abdo:2009pg,LAT:2010us}.
The well-observed extended emission shows a temporal power-law decay
up to $\simg 10^3$ s.
\item[(4)] The high-energy emission sometimes shows a large amplitude variability
on short timescales \citep{Abdo:2009pg}.
\item[(5)] The high-energy emission is delayed 
behind the onset of the MeV emission in almost all LAT GRBs
\citep{Abdo:2009nj,Abdo:2009a,Abdo:2010gr,Abdo:2009pg,LAT:2010us}.
The delay time in the rest frame is $t_{\rm delay}\sim 1$ s 
for long GRBs and $\sim 0.1$ s for short bursts, 
GRB 081024B and GRB 090510.
\item[(6)] The prompt emission spectrum of GRB 090902B 
has a quasi-blackbody component, 
which is consistent with the photospheric emission
\citep{Abdo:2009pg,Ryde:2009wn}.
\end{itemize}

The Fermi discoveries excite the theoretical reconsiderations
of the GeV emission,
which may be classified as
(i) the external shock models
with synchrotron emission
from adiabatic shocks \citep{Kumar:2009ps,Kumar:2009vx,Duran:2010et,
Corsi:2009ib,Corsi:2009vk,Gao:2009qa,DePasquale:2009bg,Pandey:2010ti}
and radiative shocks \citep{Ghirlanda:2009mj,Ghisellini:2009rw},
SSC \citep{Zou:2008dr,Wang:2009rp,Corsi:2009ib,Neamus:2010nb},
and external Compton \citep{Murase:2009su}, and
(ii) the internal shock models
with synchrotron emission \citep{Wang:2009sd,Fan:2009ua},
SSC \citep{Li:2008ub,Zou:2008dr,Abdo:2009pg,Corsi:2009ib,Corsi:2009vk},
hadronic emission \citep{Asano:2008tc,Razzaque:2009rt,Asano:2009ta},
and external Compton
of cocoon \citep{Toma:2009mw}
or photospheric emission \citep{Toma:2010xw}.
The extended GeV emission most likely has the external shock origin.
The emission mechanism could be other than synchrotron
since the maximum synchrotron cutoff
terminates the late ($\simg 100$ s) emission of $\simg 10$ GeV photons 
from external shock synchrotron
\citep{Li:2010zx,Piran:2010ew}.
On the other hand, the external shocks cannot produce the observed
large amplitude variability on short timescales
\citep{Ioka:2004gy,Sari:1997kn},
so that an additional origin, probably the internal shock emission,
is also required.
However, with the use of internal shock models, it is difficult to explain
the hard extra component that extends to the low-energy excess.
Obviously, synchrotron emission only cannot produce the extra component
in addition to the Band component.
The SSC emission usually peaks at higher energy than synchrotron
without extending to the low-energy excess.
Although hadronic models can make a low-energy excess
via direct and cascade radiation 
(e.g., synchrotron emission by secondary pairs at low energies),
the proton injection isotropic luminosity should be 
larger than $10^{55}$ erg/s, posing a challenge for these models
\citep{Asano:2009ta}.
The external Compton of cocoon \citep{Toma:2009mw}
or photospheric emission \citep{Toma:2010xw} seems viable,
but would need a fine tuning to smoothly connect
the high- and low-energy excesses that have
different emission origins in these models.
Therefore, the prompt emission,
not only the main Band component but also the extra component,
remains a mystery.

In this paper, we revisit the dissipative photosphere model
in light of the Fermi results,
scrutinizing the dissipative processes
that reproduce the spectral peak and luminosity (Yonetoku) relation
in Eq.~(\ref{eq:yonetoku}).
We suggest that the dissipation is caused by the surrounding matter
that decelerates the fireball (see \S\ref{sec:base}).
We show that the matter is loaded as
the {\it relativistic} baryon component into the fireball,
with a significant amount of energy received, 
comparable to that of the radiation component,
using the energy and momentum conservation (see \S\ref{sec:initial}).
This hitherto missing component 
can alter the fireball dynamics and spectra 
in a favorable way to solve the prompt emission.
Firstly, if not fully thermalized as expected in the baryon-poor fireball, 
the relativistic baryon component
can reexpand to a very high Lorentz factor (VHLF) $\Gamma \sim 10^3$--$10^6$
by its own relativistic collisionless pressure,
via a so-called ``collisionless bulk acceleration'' mechanism
(see \S\ref{sec:idea} for a short summary).
The subsequent VHLF internal shocks can explain
the extra high-energy component with variability in the Fermi bursts
by a single emission mechanism of synchrotron (see \S\ref{sec:spec}).
Secondly, the relativistic baryon component
could also operate as a heating source for $e^{\pm}$ 
to Comptonize the photospheric thermal photons into the observed Band spectrum
via $pp$, $p\gamma$ (Bethe-Heitler and photomeson), and Coulomb 
thermalization processes.
Without fine tuning, the relativistic baryon component
has the right amount of energy (comparable to the thermal energy)
to make the nonthermal tails (see \S\S\ref{sec:initial} and \ref{sec:MeV}).
In a sense, we consider a ``hot photosphere''
that is only partially thermalized with relativistic relic particles.
Our picture falls into
the photosphere--internal--external shock scenario \citep{Toma:2010xw},
in which the main Band emission comes from the photosphere
and the extra components come from the internal and external shocks.

This paper is organized as follows.
In \S\ref{sec:idea}, we first summarize the basic idea 
for making the VHLF fireballs.
In \S\ref{sec:dyn}, we go into the fireball dynamics, expanding the
idea of the collisionless bulk acceleration to a VHLF.
In \S\ref{sec:base}, we recall that 
the observed Yonetoku relation in Eq.~(\ref{eq:yonetoku}) strongly suggests
the fireball dissipation under the photosphere, 
probably caused by the baryon loading.
In \S\ref{sec:initial}, 
we use a simple two-body collision to describe the dissipation,
properly taking into account the relativistic hot motions
of baryons before thermalization.
We show that the relativistic baryon component
naturally achieves comparable energy to the radiation.
Then, after giving the photospheric and pionospheric radii 
in \S\ref{sec:radius},
we derive the final coasting Lorentz factor in \S\ref{sec:Gc}.
In \S\ref{sec:others}, 
we examine the thermalization processes
via $p\gamma$ (Bethe-Heitler and photomeson), Coulomb, and plasma
interactions in addition to $pp$ collisions.
We devote \S\ref{sec:check} to consistency checks
with previous works,
and \S\ref{sec:mag} to remarks on the connections
between the collisionless bulk acceleration and
the magnetic acceleration in making the VHLF fireballs.

Secondly, in \S\ref{sec:spec}, 
we apply the VHLF fireballs to the extra high-energy component
with variability in the Fermi bursts.
In \S\ref{sec:site}, we firstly argue the GRB emission site
in favor of the photosphere--internal--external shock scenario.
In \S\ref{sec:GeV}, we calculate the internal shock synchrotron spectrum
in the VHLF models,
which is found to be consistent with the observations,
and also a nice target for the future Cherenkov Telescope Array (CTA),
since the $e^{\pm}$ creation cutoff goes beyond
the TeV range in the VHLF models.
In \S\ref{sec:090926}, we discuss the possible origins of the spectral break
around $\sim 1.4$ GeV observed in the extra component of GRB 090926,
which was suggested as the $e^{\pm}$ creation cutoff for $\Gamma \sim 600$.
In the VHLF models, the spectral break could be 
the synchrotron cooling break for $\Gamma \sim 10^4$, 
or the maximum synchrotron cutoff
limited by the dynamical time, for $\Gamma \sim 10^5$.
In \S\ref{sec:delay}, we suggest that the GeV onset delay
and its weak dependence on the burst parameters can 
be naturally derived if the baryon loading at the dissipation
is rich shortly after the jet break out of the progenitor star.

Finally, in \S\ref{sec:open}, 
we discuss the future predictions and open issues.
In \S\ref{sec:neutrino}, we predict the anticorrelation of 
$\sim$TeV neutrinos and GeV $\gamma$-rays,
which might be detectable using IceCube.
In \S\ref{sec:MeV}, we suggest that
the relativistic baryon component
could transform the photospheric thermal photons 
into the observed Band spectrum.
In \S\ref{sec:yonetoku}, we go back to the origin of
the Yonetoku relation in Eq.~(\ref{eq:yonetoku}),
which implies that the baryon loading rate, i.e, 
the environmental condition, is nearly identical to any bursts.
In \S\ref{sec:jet}, we discuss 
possible configurations of the baryon loading.
We also discuss the model
implications for the steep/shallow decay of early X-ray afterglows 
in \S\ref{sec:early} and short GRBs in \S\ref{sec:short}.
We use the cgs units with $k_B=h=1$ and $Q_x=Q/10^x$,
and the standard cosmological parameters,
$\Omega_m=0.3$ and $\Omega_\Lambda=0.7$,
unless otherwise stated.

\section{Basic idea for very high Lorentz factor (VHLF)}\label{sec:idea}

In this section, we briefly summarize the essence of 
the VHLF fireball formation,
and clarify the connections between following sections.

Let us quickly recall the conventional fireball model
\citep{p86,g86,sp90,Meszaros:1999gb}.
We initially consider a fireball that is compact and radiation-dominated.
Since it is optically thick, the fireball expands by its own radiation pressure.
The Lorentz factor grows as $\Gamma \propto r$, and 
is saturated at a constant $\Gamma_c$
after almost all the radiation energy is converted into kinetic energy.
The coasting Lorentz factor $\Gamma_c$ is higher for lower baryon loads,
whereas it has an upper limit for sufficiently low baryon loads
because the fireball becomes optically thin 
in the accelerating phase before converting all the radiation energy 
into kinetic energy (i.e., before the saturation).
The maximum value of the coasting Lorentz factor 
is about $\Gamma_c \le \eta_{*} \sim 10^3$
for typical parameters [see Eq.~(\ref{eq:etast}) and \S\ref{sec:radius}].

Our idea is that even the baryon-poor fireball can 
accelerate to the saturating (very high) Lorentz factor,
if the radiation energy is transferred to the other 
relativistic component that is not radiated away
from the fireball.
In our case, this is the relativistic baryon component
(and the magnetic component resulting from it).
We initially consider an almost baryon-free fireball
(e.g., a leptonic fireball).
A small amount of baryon is loaded into the fireball
as the relativistic baryon component
from the surrounding matter near the progenitor star
or the preceding baryon-rich ejecta.
The loaded baryon decelerates the fireball,
receiving energy from the radiation in a collisionless way.
We show that half the radiation energy can be transferred to
the relativistic baryon component at the fireball dissipation
using the energy and momentum conservation
in Eqs.~(\ref{eq:econs}) and (\ref{eq:pcons}).

The fireball dissipation can occur under the photosphere
(before the fireball becomes optically thin)
so that the dissipated energy is trapped by the fireball.
The dissipation is even demanded by the GRB observations
if we identify the spectral peak energy $\varepsilon_{\rm peak}$
with the fireball temperature
(see \S\ref{sec:base} and Refs.~\cite{Rees:2004gt} and ~\cite{Thompson:2006fp}).
The observed temperature $\Gamma T'$ drops at the dissipation
because the fireball radius at the base of the flow $r_b$
effectively increases,
and this enables us to account for the observed spectral relations
in Eq.~(\ref{eq:yonetoku})
(see \S\S\ref{sec:base} and \ref{sec:yonetoku}).
The thermalization of radiation and $e^{\pm}$
is much faster than that of protons,
keeping the comoving leptonic temperature at a nonrelativistic value.

\begin{figure}
\centerline{\includegraphics[scale=.35]{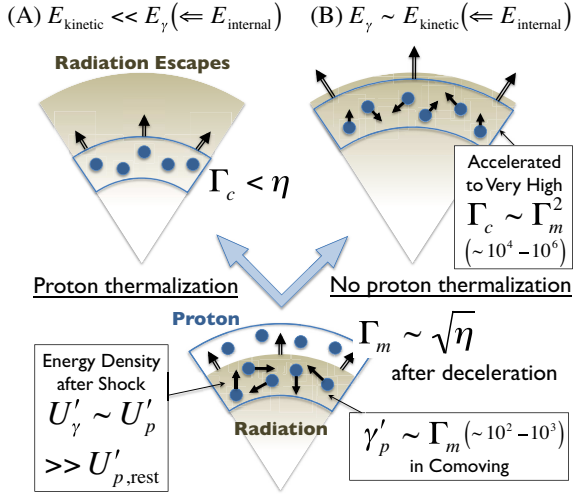}}
\caption{
Schematic of a fireball evolution with dissipation.
The radiation-dominated fireball is decelerated 
to a Lorentz factor $\Gamma_m \sim \sqrt{\eta}$ under the photosphere,
via shock with baryon (protons) initially at rest.
The comoving energy density of protons 
can be boosted to a value comparable to the radiation,
$U'_p \sim U'_\gamma$,
where almost all the proton energy is carried 
by the relativistic hot motions of protons
with a random Lorentz factor of 
$\gamma'_p \sim \Gamma_m$.
(A) If thermalization is effective,
almost all the proton energy immediately dissipates into radiation,
leading back to a standard radiation-dominated fireball.
The maximum Lorentz factor is less than $\eta_{*} \sim 10^3$
as in the conventional case.
(B) If thermalization is not effective,
the relativistic collisionless motions of protons continue to push
the fireball with $\Gamma \propto r$
up to a (saturating) VHLF,
$\Gamma_c \sim \gamma'_p \Gamma_m \sim \eta \sim 10^3$--$10^6$,
even beyond the photosphere.
The kinetic energy remains comparable to the radiation energy
since the energy density of relativistic protons 
behaves like radiation, $U'_p \sim U'_\gamma \propto r^{-4}$.
}
\label{fig:schematic}
\end{figure}

\begin{figure}
\centerline{\includegraphics{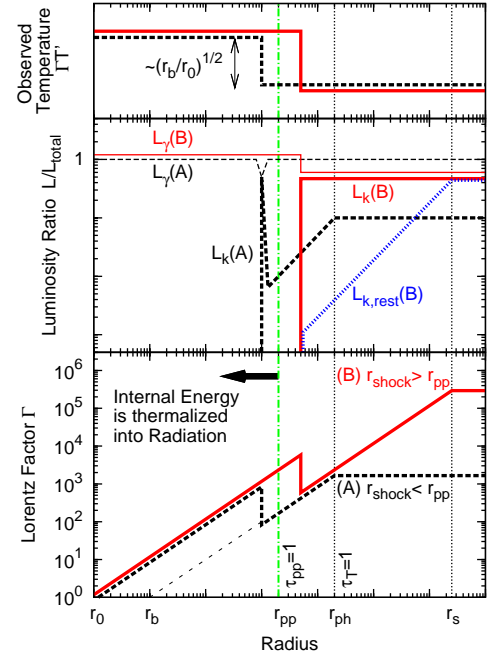}}
\caption{
Schematic evolution of physical quantities 
as a function of the fireball radius $r$.
The radiation-dominated fireball is decelerated 
via shock with baryon (protons) initially at rest,
boosting the proton energy to a value comparable to the radiation.
(A) ({\it dashed lines}) If the thermalization is effective, e.g.,
under the pionosphere $r<r_{pp}$ 
with the $pp$ collisional optical depth $\tau_{pp} > 1$,
almost all the proton kinetic luminosity $L_k(A)$
immediately dissipates into the radiation luminosity $L_\gamma(A)$.
The Lorentz factor grows under the radiation pressure,
so that it coasts at the photosphere $r=r_{\rm ph}$
with $\Gamma_c \siml 10^3$.
(B) ({\it solid lines}) If the protons are not thermalized,
the kinetic luminosity remains comparable to the radiation luminosity,
$L_k(B) \sim L_\gamma(B)$,
since almost all the proton energy is carried 
by the relativistic random motions of protons,
not by the rest mass energy, $L_{k,{\rm rest}}(B) \ll L_k(B)$.
The Lorentz factor grows under the proton collisionless pressure
up to a VHLF, $\Gamma_c \sim \Gamma_m^2 \sim 10^3$--$10^6$,
even beyond the photosphere $r>r_{\rm ph}$.
In both cases (A) and (B),
the observed temperature $\Gamma T'$
drops by $\sim (r_b/r_0)^{1/2}$ times
because the fireball radius 
at the base of the flow $r_b$ effectively increases.
}
\label{fig:concept}
\end{figure}

In previous studies, we usually (implicitly) 
assume that the baryon (proton) component
of the fireball is also completely thermalized at the dissipation.
This is the case (A) in Figs.~\ref{fig:schematic} and \ref{fig:concept},
where we schematically show the fireball evolution with dissipation.
The radiation-dominated fireball expands outward with $\Gamma \propto r$, 
and is dissipated, e.g., via shocks with baryon (protons) initially at rest.
Here, if we look closely into the dissipation,
the shock is usually collisionless at least in the early stage.
The protons are trapped by the fireball via magnetic fields,
which could be generated by the plasma instabilities
or could be advected from the central engine.
In any case, the velocity of protons is not changed so much
in the frame of the shocked region.
Then, the isotropized protons have
a random Lorentz factor of
\beqa
\gamma'_p \sim \Gamma_m\ \left(\sim 10^2-10^3\right),
\label{eq:gap}
\eeqa
in the shocked region
because the protons drive into the shocked region 
with the Lorentz factor that is about the
bulk Lorentz factor of the fireball after the merger $\Gamma_m$.
In this process,
the fireball energy is transferred to protons,
leading to an almost equipartition of the (comoving) energy density
between protons and radiation
[see \S\ref{sec:initial} and Eq.~(\ref{eq:E'm2})],
\beqa
U'_p \sim U'_\gamma,
\eeqa
according to the energy and momentum conservation
in Eqs.~(\ref{eq:econs}) and (\ref{eq:pcons}).
Thus, the kinetic luminosity temporarily equals
the radiation luminosity, $L_k(A) \sim L_{\gamma}(A)$, in Fig.~\ref{fig:concept}.
The energy equipartition also renders
the bulk Lorentz factor after the merger as
\beqa
\Gamma_m \sim \sqrt{\eta},
\eeqa
where $\eta$ is the dimensionless entropy (the radiation to baryon ratio)
of the fireball
after the merger in Eq.~(\ref{eq:etadef}).
However, almost all the proton energy is carried 
by the relativistic hot motions of protons,
not by the rest mass energy, i.e., $U'_{p,{\rm rest}} \ll U'_p$.
The subsequent thermalization of protons
via $pp$, $p\gamma$ (Bethe-Heitler and photomeson), and Coulomb interactions
(see \S\ref{sec:others})
reduces the proton energy and pressure considerably into radiation
[For example, the protons are effectively thermalized 
via $pp$ collisions under the pionosphere $r<r_{pp}$,
where the optical depth to $pp$ collisions is larger than 
unity $\tau_{pp} > 1$ (see \S\ref{sec:radius})].
Then, the fireball evolution is essentially similar to that without shocks
since the dissipated energy is trapped in the system
under the photosphere.
Therefore, the coasting Lorentz factor has a conventional upper limit of
\beqa
\Gamma_c \le \eta_{*} \sim 10^3,
\label{eq:gacB}
\eeqa
as the radiation escapes from the photosphere
before transferring its energy to the baryon kinetic energy
[see Eq.~(\ref{eq:etast}) and \S\ref{sec:radius}].

The evolution is totally different
if the thermalization is not completed.
This is the case (B) in Figs.~\ref{fig:schematic} and \ref{fig:concept}.
The evolution is almost the same as the previous case (A)
before the proton thermalization.
However, if the relativistic hot motions of protons
are not thermalized
(e.g., $pp$ collisions are not effective with $\tau_{pp} < 1$),
the relativistic collisionless motions of protons
reexpand the fireball with $\Gamma \propto r$,
acting like radiation pressure.
Since the random motions are converted into the bulk motion,
the final bulk Lorentz factor is 
the bulk Lorentz factor after the merger $\Gamma_m$ multiplied
by the random Lorentz factor $\gamma'_p \sim \Gamma_m$ as
\beqa
\Gamma_c \sim \gamma'_p \Gamma_m
\sim \Gamma_m^2\ \sim \eta\ \left(\sim 10^4-10^6\right),
\eeqa
which is a (saturating) VHLF,
much larger than the conventional upper limit $\eta_{*} \sim 10^3$ 
in Eqs.~(\ref{eq:gacB}) and (\ref{eq:etast}).
It is remarkable that
such a ``collisionless bulk acceleration'' continues beyond the photosphere,
$r>r_{\rm ph}$, i.e.,
even after radiation loses contact with matter,
in contrast with the thermalization case (A).
The kinetic luminosity also remains 
comparable to the radiation luminosity,
\beqa
L_{k}(B) \sim L_{\gamma}(B),
\eeqa
since the energy density of relativistic protons 
behaves like radiation, $U'_p \sim U'_\gamma \propto r^{-4}$.

The energy source of the collisionless bulk acceleration 
is the initial radiation energy.
The radiation energy is transferred to the relativistic baryon component
in the nonradiative form of the isotropic random hot motions,
which is later converted into the bulk kinetic energy.
Even if the relativistic baryon energy dissipates into
the magnetic field,
a similar acceleration continues via the magnetic pressure
(see \S\S\ref{sec:others} and \ref{sec:mag}).
However, the acceleration mechanism does not work 
if the baryon loading occurs after the coasting
since adding mass just reduces the bulk Lorentz factor.\footnote{
In this paper, we consider the complete merger case.
If we consider the reverse and forward shock structure,
the further acceleration is seemingly repeatable even after the coasting
since the energy can be transferred from the rear to the front shell.
However, this is not likely as
the front shell reaccelerates
before completely receiving the rear shell energy
[see \S\ref{sec:jet} and Eq.~(\ref{eq:causal})].}

According to the above considerations,
it is physically reasonable to define a VHLF as a Lorentz factor
larger than the conventional maximum value $\eta_{*} \sim 10^3$ 
in Eqs.~(\ref{eq:gacB}) and (\ref{eq:etast}).
A VHLF could open a new paradigm for interpreting the GRB properties,
removing a theoretical bias to the lower Lorentz factors.
In \S\ref{sec:spec}, we apply the VHLF fireballs
to the internal shock synchrotron model
for reproducing the keV--GeV power-law spectrum with high time variability
detected by Fermi (\S\ref{sec:GeV}), 
the spectral break in GRB 090926
by the synchrotron cooling break 
or the maximum synchrotron cutoff limited by the dynamical time
(\S\ref{sec:090926}),
and the GeV onset delay (\S\ref{sec:delay}).
In \S\ref{sec:spec}, we predict an anticorrelation between
GeV $\gamma$-rays and TeV neutrinos, and also
suggest that the relativistic baryon component
could heat the photospheric thermal photons 
into the observed Band spectrum.

Note that each element of our idea to create VHLF fireballs 
is not completely new.
The acceleration, which
converts the internal energy back into the kinetic energy,
was previously discussed 
in the context of the internal shock efficiency
\citep{Kobayashi:2001iq,Kumar:1999cv},
although their fireballs cannot reach a VHLF
since shocks occur in the coasting phase, not in the accelerating phase.
The fireball dissipation under the photosphere
was also discussed in the photosphere model
\citep{Rees:2004gt,Thompson:2006fp,Ghisellini:2007tm,Ioka:2007qk}.
The pionosphere was also discussed
for the neutrino emission and the neutron decoupling
\citep{Meszaros:2000fs,px94,dkk99,Fuller:2000nb,Beloborodov:2009be}.
However, the combination of these elements 
leads to a new concept of the VHLF fireball 
arising from the hot photosphere,
which has not been discussed so far, 
to the best of our knowledge.

It is useful to refer to 
an interesting analogy with cosmology.
The relativistic protons that are not thermalized
after the fireball dissipation
are similar to the relic particles in the Universe, 
in particular, hot relics such as
neutrinos and light dark matter, which are relativistic at the freeze out.
In this sense, a ``hot relic fireball'' attains a VHLF.
The dissipation of the GRB fireball is also 
similar to the reheating of the Universe,
both of which leads to the entropy production
after the birth of the fireball.
We use ``dissipation'' 
for both the entropy production
at the collisionless shock and
at the thermalization of relativistic protons,
and ``thermalization'' for
the dissipation of the relativistic proton energy into radiation.

\section{Fireball dynamics:
collisionless bulk acceleration to very high Lorentz factor (VHLF)}
\label{sec:dyn}

In this section, we investigate the fireball dynamics to a VHLF
via collisionless bulk acceleration in detail, 
extending the idea in the previous section.
In \S\ref{sec:base}, we first recall that the fireball dissipation
is strongly suggested by the observed spectral relation.
We suggest that the dissipation is caused by the mass loading,
not by the magnetic reconnection or neutron decay, 
for the radiation-dominated fireball.
In \S\ref{sec:initial}, 
we use a simple two-body collision to describe the dissipation,
properly taking into account the relativistic hot motions
of protons before thermalization 
to discuss the collisionless bulk acceleration.
Then, after giving the photospheric and pionospheric radii 
in \S\ref{sec:radius},
we derive the final coasting Lorentz factor in \S\ref{sec:Gc}.
In \S\ref{sec:others}, 
we examine thermalization processes 
[$p\gamma$ (Bethe-Heitler and photomeson), Coulomb, and plasma interactions]
other than $pp$ collisions,
which are relevant in some circumstances.
We devote \S\ref{sec:check} to consistency checks
with previous works,
and \S\ref{sec:mag} to remarks on the connections
between the collisionless bulk acceleration and
the magnetic acceleration in making the VHLF fireballs.

\subsection{Fireball dissipation suggested by $\varepsilon_{\rm peak}$-$L$ Yonetoku relation}\label{sec:base}

The photosphere model has many advantages
for interpreting the GRB prompt emission (see \S\ref{sec:intro}).
In this model, we identify the spectral peak energy 
$\varepsilon_{\rm peak}$ with 
the fireball photospheric temperature $T$.
The fireball is likely radiation-dominated
since the radiative efficiency is high in most GRBs.
Under these assumptions,
the fireball dissipation is strongly suggested
by the observed spectral relation, i.e.,
the $\varepsilon_{\rm peak}$-$L$ Yonetoku relation 
in Eq.~(\ref{eq:yonetoku}), as pointed out by Thompson et al. (2007)
\citep{Thompson:2006fp,Pe'er:2007xm} (see below).

In the usual picture that the engine is an 
accreting black hole or possibly a rapidly rotating magnetar,
the engine size $r_0$ is essentially constant about
a couple of Schwarzschild radii $r_0\sim 10^7$ cm
for a black hole of mass $M_{\rm BH}\sim 10 M_{\odot}$.
However, this picture ($r_0 \sim$ const) leads to
a different relation for the isotropic luminosity,
\beqa
L=4\pi r_0^2 c a T_0^4 \propto T^4,
\eeqa
from the observed Yonetoku relation $L \propto T^2$,
where the observed temperature preserves the initial temperature 
$T \sim T_0$ for radiation-dominated fireballs.
Therefore, the fireball radius is 
most likely reset by the fireball dissipation.
Note that the relation tracks $L \propto T$ after 
the fireball becomes matter-dominant.
Although we might be able to transform $L \propto T^4$ to $L \propto T^2$
by using the matter-dominant track,
this is not likely since
the low-luminosity region becomes radiatively too inefficient.

The dissipation takes place at a radius that is much larger
than the engine radius $r_0$.
We may estimate the radius of the dissipated fireball 
at the base of the flow 
by using the black body relation
$L=4 \pi (r_{\rm ph}/\Gamma_{\rm ph})^2 c a T^4$
as
\beqa
r_b \equiv \frac{r_{\rm ph}}{\Gamma_{\rm ph}} \sim 
1 \times 10^8\ {\rm cm}\ 
L_{53}^{1/2} T_{600\rm keV}^{-2}
> r_0 \sim 10^7\ {\rm cm},
\label{eq:rb}
\eeqa
where 
$r_{\rm ph}$ is the photospheric radius,
$\Gamma_{\rm ph}$ is the Lorentz factor of the radiating flow,
and we have $r_b \propto L^{-1/2}$
if we also combine the Yonetoku relation in Eq.~(\ref{eq:yonetoku}).

In general, the actual dissipation radius $r_m \sim \Gamma_m r_b$
is larger than the base radius $r_b$,
because, in order to be observed, a fireball has to have 
a relativistic bulk Lorentz factor 
$\Gamma_m$ after the dissipation (see \S\ref{sec:initial};
Note that, by definition, the Lorentz factor is unity 
as pulled back to the base of the flow with $\Gamma \propto r$).
The dissipation radius $r_m$
may be comparable to the size of the progenitor star
$r_m \sim \Gamma_m r_b \sim 10^{10}$--$10^{11}$ cm
if $\Gamma_m \sim 10^2$--$10^3$.
Actually, such dissipation is suggested by numerical simulations
as the jet interacts with the progenitor star 
\citep{Lazzati:2009xx,Zhang:2003rp,Mizuta:2004gu,Mizuta:2010gh}
(see also \S\S\ref{sec:yonetoku} and \ref{sec:jet}).

In addition, the dissipation has to be associated 
with the deceleration of the fireball.
For an impulsive dissipation like shocks,
the bulk Lorentz factor has to be decelerated by 
$\sim r_b/r_0 \sim 10\ L_{53}^{-1/2}$ times
at the dissipation,
and for a continuous dissipation,
the total change of the bulk Lorentz factor would be of the same order,
because $\Gamma \propto r$ for radiation-dominated fireballs
(see Fig.~\ref{fig:concept}).
For the fireball to be decelerated, 
the dissipation seems to be caused by
the mass loading, probably the baryon loading,
not by the magnetic reconnection or the neutron decay
(see also \S\S\ref{sec:yonetoku} and \ref{sec:jet}).

\subsection{Initial condition of dissipated fireballs}\label{sec:initial}

As we have discussed in the previous section,
the $\varepsilon_{\rm peak}$-$L$ Yonetoku relation suggests 
the fireball dissipation under the photosphere
due to the baryon loading.
The main features of the dissipation can be
described by a simple two-body collision.
In contrast to the previous calculations,
we approximately take into account the relativistic hot motions
of protons to discuss the collisionless bulk acceleration.

We consider a rapid shell that is radiation-dominated
with internal energy $E'_r\ (\propto r^{-1})$ 
and Lorentz factor $\Gamma_r\ (\propto r) \gg 1$,
merging with a slow mass $M_s$ with a Lorentz factor $\Gamma_s$.
Here, $\Gamma_s=1$ for the matter located near the progenitor star,
and $\Gamma_s>1$ for internal shocks.
The total energy in the lab frame is also rewritten as
\beqa
E'_r \Gamma_r = L_j t_v,
\label{eq:ltv}
\eeqa
using the duration $t_v$ and
the geometry-corrected jet luminosity $L_j=L (\theta_j/2)^2$.
We assume no baryon in the rapid shell for simplicity.
Although radiation may be already decoupled from the 
baryon-poor rapid shell before the merger,
the radiation is trapped again by the merged shell
if the radius is under the photosphere.

The energy and momentum conservation gives
\beqa
E'_r \Gamma_r + M_s c^2 \Gamma_s &=& 
\left(\Gamma_{ms} M_s c^2 + E'_m\right) \Gamma_m,
\label{eq:econs}
\\
E'_r \sqrt{\Gamma_r^2 - 1} + M_s c^2 \sqrt{\Gamma_s^2 - 1} &=& 
\left(\Gamma_{ms} M_s c^2 + E'_m\right) \sqrt{\Gamma_m^2-1},
\label{eq:pcons}
\eeqa
where $E'_m$ and $\Gamma_m$ are the internal energy and
the bulk Lorentz factor after the merger, respectively, and
\beqa
\Gamma_{ms}=\Gamma_m \Gamma_s - \sqrt{\Gamma_m^2-1} \sqrt{\Gamma_s^2-1}
\eeqa
is the relative Lorentz factor between $\Gamma_m$ and $\Gamma_s$.
The point different from the previous studies is that
we take into account the prethermalized relativistic motions
of protons with $\Gamma_{ms}$
in Eqs.~(\ref{eq:econs}) and (\ref{eq:pcons}).
Of course, this treatment is approximate but 
valid for order-of-magnitude estimates.
If we consider in the rest frame of the shocked region,
the protons run into the shocked region with $\sim \Gamma_{ms}$
and are isotropized by magnetic fields.
Here, weak magnetization is sufficient to trap protons,
and the magnetic fields could be generated by the plasma instabilities 
or could be advected from the central engine.
Since the shock is collisionless for protons before thermalization,
the proton velocities are not altered so much in the frame of the
shocked region.
Therefore, the random Lorentz factor of protons would also be
about $\Gamma_{ms}$ in the shocked region.
Before the complete deceleration,
the relative Lorentz factor between the preshocked protons
and the shocked region is larger than $\Gamma_{ms}$.
Thus, a fraction of protons would have 
the random Lorentz factor larger than $\Gamma_{ms}$.
However, at least half of the protons are shocked
after the shocked region is well decelerated,
thereby having the random Lorentz factor of $\sim \Gamma_{ms}$.
We do not consider the particle acceleration 
at the shock for simplicity.

We can solve two equations, Eqs.~(\ref{eq:econs}) and (\ref{eq:pcons}), for
two unknowns, $\Gamma_m$ and $E'_m$, as
\beqa
\Gamma_m &=& \frac{E'_r \Gamma_r + M_s c^2 \Gamma_s}
{\sqrt{{E'_r}^2 + M_s^2 c^4 + 2 E'_r M_s c^2 \Gamma_{rs}}},
\label{eq:Gm}
\\
E'_m&=&\sqrt{{E'_r}^2 + M_s^2 c^4 + 2 E'_r M_s c^2 \Gamma_{rs}}
- \Gamma_{ms} M_s c^2,
\label{eq:E'm}
\eeqa
where 
$\Gamma_{rs}=\Gamma_r \Gamma_s - \sqrt{\Gamma_r^2-1} \sqrt{\Gamma_s^2-1}$
is the relative Lorentz factor between rapid and slow shells,
and $\Gamma_{rs} \sim \Gamma_r/2 \Gamma_s$ for $\Gamma_r \gg 1$.

We are now considering the following case:
\begin{itemize}
\item[(i)] the energy is dominated by the rapid shell,
$E'_r \Gamma_r \gg M_s c^2 \Gamma_s$, so that
$E'_r \Gamma_{rs} \gg M_s c^2$ and also
the merged shell is still relativistic
$\Gamma_m \gg 1$, i.e., $\Gamma_{ms} \sim \Gamma_m/2\Gamma_s$,

\item[(ii)] the rapid shell decelerates effectively
$\Gamma_m < \Gamma_r/2$, converting the kinetic energy
into the internal energy, so that
$2 M_s c^2 \Gamma_{rs} \gg E'_r$.
\end{itemize}
That is, the slow mass is in the range,
\beqa
10^{-11} M_{\odot}\ L_{53} \theta_j^2 t_{v,-3} \Gamma_{r,3}^{-2} \Gamma_s^2
\ll M_s \Gamma_s \ll 10^{-5}\ M_{\odot}\ L_{53} \theta_j^2 t_{v,-3},
\eeqa
with Eq.~(\ref{eq:ltv}),
and thereby, only a small fraction of the progenitor mass is sufficient
to decelerate the fireball efficiently.
Then, we can simplify Eqs.~(\ref{eq:Gm}) and (\ref{eq:E'm}) as
\beqa
\Gamma_m &\sim& \left(\frac{E'_r \Gamma_r \Gamma_s}{M_s c^2}\right)^{1/2}
= \sqrt{\Gamma_s \eta}\ ,
\label{eq:Gm2}
\\
E'_m &\sim& \Gamma_{ms} M_s c^2,
\label{eq:E'm2}
\eeqa
where we use Eq.~(\ref{eq:ltv}) in the last equality 
in Eq.~(\ref{eq:Gm2}), and introduce
a dimensionless entropy (the radiation-to-baryon ratio) 
of the fireball after the merger as
\beqa
\eta \equiv \frac{E'_r \Gamma_r}{M_s c^2}=
\frac{L_j t_v}{M_s c^2} \equiv \frac{L}{\dot{M} c^2}.
\label{eq:etadef}
\eeqa
Equations~(\ref{eq:Gm2}) and (\ref{eq:E'm2})
have important implications for the initial condition of 
the dissipated fireballs:
\begin{itemize}
\item First, we can interpret Eq.~(\ref{eq:E'm2}) that
the proton energy is boosted to a value comparable to the radiation $E'_m$
and almost all the proton energy $\Gamma_{ms} M_s c^2$
is carried by their relativistic hot motions with 
the Lorentz factor
\beqa
\gamma'_p \sim \Gamma_{ms} \sim \frac{\Gamma_m}{2\Gamma_s}.
\label{eq:g'p}
\eeqa

\item Second, if the relativistic hot motions are converted 
into the bulk motion,
the final bulk Lorentz factor achieves the saturation level,
\beqa
\Gamma_{c} \sim \Gamma_m \gamma'_{p}
\sim \frac{\Gamma_m^2}{2\Gamma_s} \sim \frac{\eta}{2},
\label{eq:Gc}
\eeqa
with Eqs.~(\ref{eq:Gm2}) and (\ref{eq:g'p}),
so that the kinetic luminosity becomes comparable to the total luminosity,
$\Gamma_c \dot M c^2 \sim L/2$.

\item Finally, the fireballs should be relativistic
even after the dissipation in order to produce the observed GRBs;
otherwise, the final Lorentz factor cannot be sufficiently high,
$\Gamma_c > 10^2$--$10^3$,
to avoid the compactness problem.
\end{itemize}

Note that the bulk Lorentz factor after the merger $\Gamma_m$ 
is the same as Eq.~(\ref{eq:Gm2}) even if we omit the new factor $\Gamma_{ms}$ 
in Eqs.~(\ref{eq:econs}) and (\ref{eq:pcons})
as in the conventional case.

\subsection{Photosphere and pionosphere of fireballs}\label{sec:radius}

In this section, we summarize several important radii
of the fireballs, 
in particular, for the photosphere and pionosphere
\citep{Rees:2004gt,Meszaros:1999gb,Meszaros:2002vh}
to derive the final coasting Lorentz factor in the next section.
We assume that the fireballs are created intermittently with 
a variability timescale $t_v \simeq r_b/c$, 
which is determined by the size at the base of the flow,
for simplicity,
although we may apply the following arguments as long as the timescale
is longer than the causal timescale, $t_v > r/c \Gamma^2$.

Baryonic photosphere $r_{\rm ph}$ is determined by $\tau_T=1$,
where the Thompson optical depth to 
electrons associated with protons,
$\tau_T = n'_p \sigma_T r/\Gamma$, is unity.\footnote
{In Ref.~{\cite{Meszaros:2002vh}},
$\tau_T = n' \sigma_T \Delta'$ was used 
in the discrete shell regime.
However, we think that $\tau_T = n' \sigma_T r/\Gamma$ is correct
since photons can travel only a distance $\sim r/\Gamma$ 
within the comoving time at radius $r$.}
The Lorentz factor is $\Gamma=[r/r_b,\eta]$,
the comoving width of the shell is $\Delta' = [r, r_b \eta]$,
and the comoving volume of the shell is
$V'=4\pi r^2 \Delta'=[4\pi r^3, 4\pi \eta r_b r^2]$
when the radius $r$ is $[<r_b \eta, > r_b \eta]$.
Then, with the baryon density $n'_p=L t_v/\eta m_p c^2 V'$,
we can derive the photospheric radius as
\beqa
\frac{r_{\rm ph}}{r_b}=
\left\{
\begin{array}{ll}
\eta_{*}^{4/3} \eta^{-1/3} & \quad {\rm for} \quad \eta > \eta_{*}\\
\eta_{*}^{4} \eta^{-3} & \quad {\rm for} \quad \eta < \eta_{*},
\end{array}\right.
\label{eq:rph}
\eeqa
where 
\beqa
\eta_{*}=\left(\frac{L \sigma_T}
{4\pi m_p c^3 r_{b}}\right)^{1/4}
\simeq 1\times 10^{3} L_{53}^{1/4} r_{b,8}^{-1/4}
\label{eq:etast}
\eeqa
is a critical dimensionless entropy.
We can apply the above relations to the wind regime,
$r \simg c t_v \eta^2 \simeq r_b \eta^2$,
where 
the successive shells expand their thickness 
and overlap through internal shocks,
because $n'_p=L/4\pi r^2 m_p c^3 \eta \Gamma$
and the Lorentz factor is saturated at $r=r_b \eta$ with $\Gamma=\eta$.

Pionosphere is defined by $\tau_{pp}=1$,
where the optical depth to $pp$ collisions 
(i.e., pionic optical depth) is unity.
Repeating the previous calculations, we have
\beqa
\frac{r_{pp}}{r_b}=
\left\{
\begin{array}{ll}
\eta_{pp}^{4/3} \eta^{-1/3} & \quad {\rm for} \quad \eta > \eta_{pp}\\
\eta_{pp}^{4} \eta^{-3} & \quad {\rm for} \quad \eta < \eta_{pp},
\end{array}\right.
\label{eq:rpp}
\eeqa
where 
\beqa
\eta_{pp}=\left(\frac{L \sigma_{pp}}
{4\pi m_p c^3 r_{b}}\right)^{1/4}
=\left(\frac{\sigma_{pp}}{\sigma_T}\right)^{1/4} \eta_{*}
\simeq 500 L_{53}^{1/4} r_{b,8}^{-1/4}.
\eeqa

An $e^{\pm}$ photosphere can be formed
beyond the baryonic photosphere 
\citep{Pilla:1997jm,Guetta:2000ye,Kobayashi:2001ve,
Meszaros:2002vh,Li:2003hq,Rees:2004gt,Ioka:2007qk}.
Although there are some uncertainties 
in the amount of $e^{\pm}$,
the actual $e^{\pm}$ abundance will be between the following three cases:

(1) The $e^{\pm}$-$p$ equal mass case,
in which the rest mass energy density of $e^{\pm}$ is equal to that of baryon.
We can obtain the $e^{\pm}$ photospheric radius $r_{\pm}$
by replacing $m_p$ and $m_e$ in the baryonic photosphere case as
\beqa
\frac{r_{\pm}}{r_b}=
\left\{
\begin{array}{ll}
\eta_{\pm}^{4/3} \eta^{-1/3} & \quad {\rm for} \quad \eta > \eta_{\pm}\\
\eta_{\pm}^{4} \eta^{-3} & \quad {\rm for} \quad \eta < \eta_{\pm},
\end{array}\right.
\label{eq:rpm}
\eeqa
where
\beqa
\eta_{\pm}=\left(\frac{L \sigma_{T}}
{4\pi m_e c^3 r_{b}}\right)^{1/4}
=\left(\frac{m_p}{m_e}\right)^{1/4} \eta_{*}
\simeq 7 \times 10^3 L_{53}^{1/4} r_{b,8}^{-1/4}.
\eeqa

(2) The feasible case, in which
the $e^{\pm}$ density is determined by 
the balance between $e^{\pm}$ annihilation and creation, 
where $e^{\pm}$ is created by $pp$ collisions.
This is guaranteed
in the presence of relativistic hot protons,
which remains not thermalized after the fireball dissipation.

The inelastic cross section for $pp$ collisions is about
$\sigma_{pp} \sim 3 \times 10^{-26}$ cm$^{-2}$ 
above the pion production threshold $\sim 140$ MeV.
The inelasticity $K_{pp}$ is $\sim 0.5$, so only a few collisions are required
to extract most of the energy of the primary particles.
The energy is initially given to produce $\pi^{\pm}$ and $\pi^{0}$,
where the $\pi$ multiplicity is typically 
${\cal M}_\pi \sim 1$--$3$ near the threshold
$\sqrt{s} \sim 1$ GeV and
weakly depends on the center-of-mass energy as ${\cal M}_{\pi} \propto \ln \sqrt{s}$
\citep{GrosseOetringhaus:2009kz}.
The pions immediately decay as
$\pi^{+} \to \mu^{+} + \nu_{\mu} \to e^{+} + \nu_e + {\bar \nu}_{\mu} + \nu_{\mu}$,
$\pi^{-} \to \mu^{-} + {\bar \nu}_{\mu} \to e^{-} + {\bar \nu}_e + \nu_{\mu} + {\bar \nu}_{\mu}$,
and $\pi^0 \to \gamma + \gamma$,
and the decay gamma-rays create $e^\pm$ via $\gamma \gamma$ interactions.
Thus, the minimum $e^{\pm}$ multiplicity 
without considering the following cascade is
\beqa
{\cal M}_{\pm}^{\min} \sim 2 {\cal M}_{\pi}.
\eeqa
Alternatively, since the injected $e^{\pm}$ has a large Lorentz factor
$\gamma'_{\pm} \sim K_{pp} \gamma'_p m_p/2 {\cal M}_{\pi} m_e$,
the electromagnetic cascade will follow
and could achieve the maximum $e^{\pm}$ multiplicity,
\beqa
{\cal M}_{\pm}^{\max} \sim K_{pp} \gamma'_p m_p/m_e,
\eeqa
where $\gamma'_p$ is the comoving Lorentz factor of protons.
The minimum $e^{\pm}$ multiplicity would be appropriate
when the synchrotron cooling dominates,
since the synchrotron photons
are typically soft $\siml 1$ MeV,
while the maximum $e^{\pm}$ multiplicity
would be valid when the inverse Compton cooling dominates,
since the relativistic $e^{\pm}$ can scatter photons to high-energy
$\gg 1$ MeV.
Hereafter, we parametrize the $e^{\pm}$ multiplicity
as 
\beqa
{\cal M}_{\pm} = f_{\pm} \gamma'_p m_p/m_e.
\label{eq:fpm}
\eeqa
Then, equating the annihilation rate
\beqa
{\dot n}'_{\pm} = \frac{3}{8} n'_{+} n'_{-} \sigma_T c 
\eeqa
with the creation rate
\beqa
{\dot n}'_{\pm}
= {\cal M}_{\pm} {n'}_p^2 \sigma_{pp} c,
\eeqa
we obtain the feasible $e^{\pm}$ density $n_{\pm} = n_{+} \sim n_{-}$ as
\beqa
\frac{n'_{\pm}}{n'_p} = 
\left(\frac{8}{3} \frac{\sigma_{pp}}{\sigma_T} 
\frac{m_p}{m_e} f_{\pm} \gamma'_p \right)^{1/2},
\label{eq:emin}
\eeqa
which can be much larger than unity.

(3) The maximum case,
in which the comoving radiation energy is almost converted to 
the $e^{\pm}$ rest mass energy.
This extreme case might be realized if
the radiation spectrum has a significant fraction of
energy above the pair production threshold 
and the $e^{\pm}$ pairs develop an electromagnetic cascade.
With the $e^{\pm}$ density $n'_{\pm,\max}=L t_v/\Gamma m_e c^2 V'$,
the optical depth condition 
$\tau_T=n'_{\pm,\max} \sigma_T r/\Gamma=1$ gives
\beqa
\frac{r_{\pm,\max}}{r_b}=
\left\{
\begin{array}{ll}
\eta_{\pm} & \quad {\rm for} \quad \eta > {\eta_{\pm}}\\
\eta_{\pm}^{4} \eta^{-3} & \quad {\rm for} \quad \eta < {\eta_{\pm}}.
\end{array}\right.
\label{eq:rpmmax}
\eeqa
The $e^{\pm}$ rest mass energy surpasses the baryonic one,
because the proton energy, 
comparable to the radiation energy in Eq.~(\ref{eq:E'm2}),
is carried by relativistic motions, not by the rest mass energy.
The acceleration terminates at this radius $r_{\pm,\max}$
since the $e^{\pm}$ rest mass energy almost equals the total energy.

\subsection{Final coasting Lorentz factor of dissipated fireballs}
\label{sec:Gc}

\begin{figure}
\centerline{\includegraphics{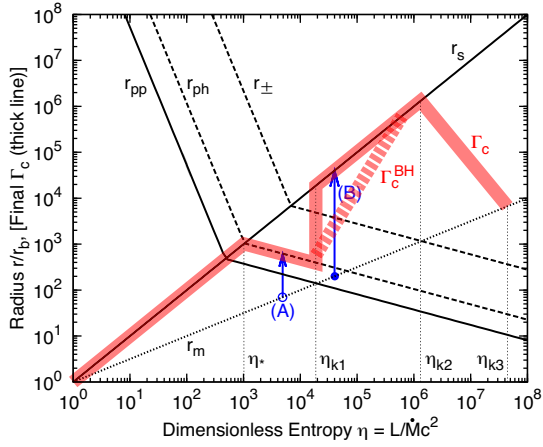}}
\caption{
$\eta$--$r$ diagram to read normalized radii $r/r_b$ and 
the final coasting Lorentz factor $\Gamma_c$ in Eq.~(\ref{eq:Gcall})
({\it red thick line})
and in Eq.~(\ref{eq:GcBH})
(with efficient Bethe-Heitler processes; {\it red thick dotted line})
as a function of the dimensionless entropy $\eta=L/{\dot M} c^2$
(the radiation to baryon ratio).
A fireball can achieve a VHLF, $\Gamma_c > \eta_{*}$,
if the dissipation radius $r_m$ in Eq.~(\ref{eq:rm})
exceeds the pionosphere $r_{pp}$ 
in Eq.~(\ref{eq:rph}) (pionic optical depth $\tau_{pp}=1$).
The maximum coasting Lorentz factor reaches $\Gamma_{c,\max} \sim 10^6$ 
in Eqs.~(\ref{eq:Gcmax1}) and (\ref{eq:Gcmax2}).
We also show the baryonic photosphere $r_{\rm ph}$ in Eq.~(\ref{eq:rpp}) 
(Thompson optical depth $\tau_{T}=1$),
the $e^{\pm}$ photosphere $r_{\pm}$ in Eq.~(\ref{eq:rpm}) 
for the $e^{\pm}$-$p$ equal mass case,
and the saturation radius $r_s/r_b=\eta$.
Critical entropies $\eta_{*}$, $\eta_{k1}$, $\eta_{k2}$, and $\eta_{k3}$
are given by Eqs.~(\ref{eq:etast}), (\ref{eq:etak1}), (\ref{eq:etak2}),
and (\ref{eq:etak3}),
respectively.
}
\label{fig:etar}
\end{figure}

Figure~\ref{fig:etar} shows the $\eta$--$r$ diagram,
\citep{Rees:1994nw,Meszaros:1999gb,Meszaros:2000fs}
which is useful to read out the final coasting Lorentz factor.
In the conventional picture,
we usually start the fireball evolution from the base of the flow, $r/r_b=1$,
for a given entropy $\eta$.
The fireball expands with $\Gamma \propto r$ 
as far as below the saturation radius $r_s$
and the photospheric radius $r_{\rm ph}$.
Then, the maximum coasting Lorentz factor is $\sim \eta_{*}$
in Eq.~(\ref{eq:etast}).

However, the starting radius is different from the base, $r/r_b=1$,
for the dissipated fireball case
because the observed fireballs have to be relativistic after dissipation,
as discussed in \S\ref{sec:initial}.
Then, using $\Gamma \propto r$,
we can plot the starting radius $r_m$ in the diagram 
({\it dotted line} in Fig.~\ref{fig:etar}) as
\beqa
\frac{r_m}{r_b} \sim \Gamma_m \sim \sqrt{\Gamma_s \eta},
\label{eq:rm}
\eeqa
where $\Gamma_m$ is the bulk Lorentz factor after dissipation
in Eq.~(\ref{eq:Gm2}),
derived from the energy and momentum conservation
in Eqs.~(\ref{eq:econs}) and (\ref{eq:pcons}),
and Fig.~\ref{fig:etar} shows the case $\Gamma_s=1$.
Here, we take the independent model parameters as 
the dimensionless entropy $\eta$ in Eq.~(\ref{eq:etadef}),
the Lorentz factor of the slow mass $\Gamma_s$, 
the total isotropic luminosity $L$, 
and the engine size $r_0\ (<r_b)$ in \S\ref{sec:base}.
We can derive the other parameters, such as
the observed temperature $T$ in Eq.~(\ref{eq:yonetoku}),
the radius at the base of the flow
$r_b$ in Eq.~(\ref{eq:rb}),
the dissipation radius $r_m$ in Eq.~(\ref{eq:rm}),
and the Lorentz factor of the rapid shell before merger
$\Gamma_r=r_m/r_0$.

As long as the dissipation radius is below the pionosphere,
$r_m < r_{pp}$, i.e., in the low to moderate entropy range $\eta < \eta_{k1}$ 
(see Fig.~\ref{fig:etar}) where
\beqa
\eta_{k1}=\Gamma_s^{-3/5} \eta_{pp}^{8/5}
\simeq 2 \times 10^4\ \Gamma_s^{-3/5} L_{53}^{2/5} r_{b,8}^{-2/5},
\label{eq:etak1}
\eeqa
with Eqs.~(\ref{eq:rpp}) and (\ref{eq:rm}),
the fireball evolution is essentially similar to the conventional case.
This is the case (A) in Fig.~\ref{fig:etar}.
The relativistic hot motions of protons, acquired at the 
merger with a radiation-dominated fireball,
are quickly thermalized into radiation 
via $pp$ collisions under the pionosphere.
The dissipated fireball goes back to the standard radiation-dominated fireball,
and hence, the maximum coasting Lorentz factor is again $\sim \eta_{*}$
in Eq.~(\ref{eq:etast}), 
as the radiation escapes from the photosphere $r=r_{\rm ph}$.
Such a fireball is sometimes called a reborn fireball \citep{Ghisellini:2007tm}.

The evolution is completely different
if the dissipation radius exceeds the pionosphere
(see \S\ref{sec:others} for other thermalization processes).
This is the case (B) in Fig.~\ref{fig:etar}.
Relativistic hot motions of protons
are not effectively thermalized,
which are converted into the bulk motion,
leading to a VHLF up to the saturation value
$\Gamma_c \sim \eta$ in Eq.~(\ref{eq:Gc}).
Here, the acceleration $\Gamma \propto r$ continues even beyond the photosphere
because the pressure is provided by the collisionless hot motions of protons.
For a large dissipation radius exceeding
the baryonic photosphere $r_m>r_{\rm ph}$,
the $e^{\pm}$ creation is necessary 
to trap the radiation that boosts the proton energy.
Although the $e^{\pm}$ abundance is somewhat uncertain
(see \S\ref{sec:radius}),
we first derive the upper limits on the coasting Lorentz factor
by considering the most favorable case.
For a dissipation radius $r_m>r_{\pm}$ in Eq.~(\ref{eq:rpm}),
i.e., in the range $\eta > \eta_{k2}$ where
\beqa
\eta_{k2}=\Gamma_s^{-3/5} \eta_{\pm}^{8/5}
\simeq 1 \times 10^6\ \Gamma_s^{-3/5} L_{53}^{2/5} r_{b,8}^{-2/5},
\label{eq:etak2}
\eeqa
the $e^{\pm}$ rest mass necessary for trapping the radiation
dominates the baryon rest mass,
suppressing the coasting Lorentz factor below the saturation value $\eta$
in Fig.~\ref{fig:etar}.
An upper limit to the dissipation radius is
$r_m<r_{\pm,\max}$ in Eq.~(\ref{eq:rpmmax}), 
i.e., in the range $\eta < \eta_{k3}$ where
\beqa
\eta_{k3}=\Gamma_s^{-1} \eta_{\pm}^{2}
\simeq 4 \times 10^7\ \Gamma_s^{-1} L_{53}^{1/2} r_{b,8}^{-1/2},
\label{eq:etak3}
\eeqa
because the fireball larger than $r_{\pm,\max}$
is too rarefied to trap the radiation even if $e^{\pm}$ is maximally created.
Thus, a dissipated fireball with $\eta > \eta_{k3}$
is never formed via physical processes.
Since the $e^{\pm}$ rest mass equals the total energy at 
$r_m=r_{\pm,\max}$ in Eq.~(\ref{eq:rpmmax}),
the dissipated fireball terminates its acceleration.

The final coasting Lorentz factor 
({\it red thick line} in Fig.~\ref{fig:etar})
may be summarized as
\beqa
\Gamma_c=
\left\{
\begin{array}{ll}
\eta & \quad {\rm for} \quad \eta < \eta_{*}\\
\eta_{*}^{4/3} \eta^{-1/3} & \quad {\rm for} \quad \eta_{*} < \eta < \eta_{k1}\\
\eta & \quad {\rm for} \quad \eta_{k1} < \eta < \eta_{k2}\\
\eta_{k2} (\eta/\eta_{k2})^{\lambda} & \quad {\rm for} 
\quad \eta_{k2} < \eta < \eta_{k3},
\end{array}\right.
\label{eq:Gcall}
\eeqa
where the index $\lambda$ is determined by
the condition
$\Gamma_c=r_m/r_b=\sqrt{\Gamma_s \eta_{k3}}$ at $\eta=\eta_{k3}$
(see Fig.~\ref{fig:etar})
as
\beqa
\lambda=-\frac{\ln \left(\eta_{k2} \Gamma_s^{-1/2} \eta_{k3}^{-1/2}\right)}{\ln \left(\eta_{k3} \eta_{k2}^{-1}\right)}.
\eeqa
The appropriate coasting Lorentz factor for $\eta_{*} < \eta < \eta_{k1}$
would be $\eta_{*}$ instead of the slightly smaller $\eta_{*}^{4/3} \eta^{-1/3}$
in Eq.~(\ref{eq:Gcall})
because most of the electrons above the photosphere
can still scatter with a decreasing fraction of free-streaming photons,
and continue accelerating as long as the comoving Compton drag time
$t'_{\rm drag}=m_p c^2/c \sigma_T U'_{\gamma}$
is less than the comoving dynamical time
$t'_{\rm dyn}=r/c\Gamma$
\citep{Grimsrud:1998me,Meszaros:1999gb}.
The ratio of these two times,
$t'_{\rm drag}/t'_{\rm dyn} = 4\pi m_p c^3 r \Gamma^3/L \sigma_T
=(r/\eta_* r_b)^4$,
gives the coasting radius where the radiative acceleration ceases
at $r/r_b=\eta_* > r_{\rm ph}/r_b$
for $\eta > \eta_*$.

The maximum value of the coasting Lorentz factor is
\beqa
\Gamma_{c,\max}=\eta_{k2}
\simeq 1 \times 10^6\ \Gamma_s^{-3/5} L_{53}^{2/5} r_{b,8}^{-2/5},
\label{eq:Gcmax1}
\eeqa
(see Fig.~\ref{fig:etar}),
which is a VHLF, much larger than the conventional upper limit 
$\eta_{*} \sim 10^3$
in Eqs.~(\ref{eq:gacB}) and (\ref{eq:etast}).
This maximum value is realized by the $e^{\pm}$-$p$ equal mass case 
when the $e^{\pm}$ rest mass energy is equal to the baryonic one.
If the actual $e^{\pm}$ abundance is less than this case,
the radiation escapes without completely transferring its energy to
the proton component, leading to a smaller $\Gamma_{c,\max}$.
A more conservative estimate of the maximum Lorentz factor
is provided by the feasible case of the $e^{\pm}$ creation
in Eq.~(\ref{eq:emin}) of \S\ref{sec:radius},
which gives the feasible optical depth to trap the radiation at the dissipation.
By solving $\tau_T=2 n'_{\pm} \sigma_T r_m/\Gamma_m=1$ 
in terms of $\eta$, with Eqs.~(\ref{eq:emin}),
(\ref{eq:rm}), (\ref{eq:etast}), (\ref{eq:g'p}), and
$n_p'=L/4\pi r_m^2 m_p c^3 \eta \Gamma_m$,
we may identify its solution with the most conservative maximum Lorentz factor 
as
\beqa
\Gamma_{c,\max}^{c}
&=&\left(
\frac{16}{3}
\frac{\sigma_{pp}}{\sigma_T}
\frac{m_p}{m_e}
f_{\pm}\right)^{2/9}
\Gamma_s^{-7/9}
\eta_{*}^{16/9}
\nonumber\\
&\sim& 8 \times 10^5
f_{\pm}^{2/9} \Gamma_s^{-7/9} L_{53}^{4/9} r_{b,8}^{-4/9},
\label{eq:Gcmax2}
\eeqa
which is still a VHLF unless $f_{\pm}$ is very small in Eq.~(\ref{eq:fpm}).
Therefore, we conclude that
the coasting Lorentz factor can attain a VHLF,
$\Gamma_c \sim 10^4$--$10^6$,
if the dissipation radius is above the pionosphere, i.e.,
in the high entropy range
with $\eta > \eta_{k1} \sim 10^4 \Gamma_s^{-3/5}$
(see \S\ref{sec:others} for other thermalization processes).
It is interesting to note a sharp rise of $\Gamma_{c}$
at $\eta=\eta_{k1}$, where
a slight change of the baryon loads could lead to
a large difference in the coasting Lorentz factor
$\Delta \Gamma_c/\Gamma_c > 10$.

The initial fireball energy is shared by
the photospheric radiation luminosity $L_{\rm ph}$,
the neutrino luminosity $L_{\nu}$ (see \S\ref{sec:neutrino}),
and the kinetic luminosity of protons and $e^{\pm}$,
$L_k = L_p + L_{\pm}$,
where the kinetic luminosity
is later dissipated via internal and external shocks
(see \S\ref{sec:site}).
Table~\ref{tab:lumi} summarizes the final shares
of the initial total luminosity $L$
as a function of $\eta$.
As we discuss in \S\ref{sec:site}, if we connect $L_{\rm ph}$ 
with the Band spectral component
and $L_k$ with the extra high-energy power-law component,
the equal contributions from both components
could suggest the VHLF range with $\eta_{k1} < \eta$
in Table~\ref{tab:lumi}, Fig.~\ref{fig:etar} and Eq.~(\ref{eq:Gcall}),
and the GeV onset delay phase
could suggest the moderate entropy range $\eta_{*} < \eta < \eta_{k1}$
(see \S\ref{sec:delay}).

\begin{widetext}
\begin{center}
\begin{deluxetable}{cccccc}
\tablecolumns{6}
\tablecaption{
Final shares of the initial total luminosity $L$
as a function of the radiation-to-baryon ratio $\eta$.
The luminosity is shared by the photospheric radiation $L_{\rm ph}$,
neutrinos $L_{\nu}$ (see \S\ref{sec:neutrino}),
and protons and $e^{\pm}$ in the kinetic form
$L_k = L_p + L_{\pm}$,
which is later dissipated via internal/external shocks
(see \S\ref{sec:site}).
In \S\ref{sec:site}, we connect $L_{\rm ph}$ with the Band spectral component
and $L_k$ with the extra high-energy power-law component (PL in short),
respectively.
Note that the coasting Lorentz factor $\Gamma_c$
achieves a VHLF in the high entropy range $\eta_{k1} < \eta$
[See Fig.~\ref{fig:etar} and Eq.~(\ref{eq:Gcall})].
Critical entropies $\eta_{*}$, $\eta_{k1}$, $\eta_{k2}$, and $\eta_{k3}$
are given by Eqs.~(\ref{eq:etast}), (\ref{eq:etak1}), (\ref{eq:etak2}),
and (\ref{eq:etak3}),
respectively.
\label{tab:lumi}
}
\tablehead{ $\eta$ & $\Gamma_c$ & $L_{\rm ph}$ [$\sim$Band] & $L_{k}$ [$\sim$PL] & Spectrum & $L_{\nu}$ }
\startdata
$1<\eta<\eta_{*}\sim 10^3$ & $\eta$ & $\ll L$ & $\sim L$ & PL & $\sim L$ \nl
$\eta_{*}<\eta<\eta_{k1}\sim 10^4$ & $\eta_*^{4/3} \eta^{-1/3}$ (or $\eta_*$) & $\sim L$ & $\ll L$ & Band & $\sim L$ \nl
$\eta_{k1}<\eta<\eta_{k2} \sim 10^6$ & $\eta$ & $\sim L$ & $\sim L_{p} \sim L$ & Band+PL & $\ll L$ \nl
$\eta_{k2}<\eta<\eta_{k3} \sim 10^7$ & $\eta_{k2} (\eta/\eta_{k2})^{\lambda}$ & $\sim L$ & $\sim L_{\pm} \sim L$ & Band+PL & $\ll L$ \nl
\enddata
\end{deluxetable}
\end{center}
\end{widetext}

\subsection{Thermalization: $pp$, $p\gamma$, Coulomb,
and plasma processes}
\label{sec:others}

We have considered $pp$ collisions for the thermalization process
of relativistic protons that are entrained in the fireball dissipation.
In this section, we also examine other processes,
$p\gamma$ (Bethe-Heitler and photomeson) and Coulomb interactions,
which are relevant in some circumstances.
(The bremsstrahlung emission is not effective for fiducial parameters.)

The main competing interaction is Bethe-Heitler photopair production
($p\gamma \to p e^{+} e^{-}$).
The threshold energy of photons normalized by
the peak energy is about
\beqa
\frac{2 m_e c^2/\gamma'_p}{\varepsilon'_{\rm peak}}
\sim \frac{2 m_e c^2}{\varepsilon_{\rm peak}} 
2 \Gamma_s \left(\frac{r}{r_m}\right)^2
\sim 3\ L_{53}^{-1/2} \Gamma_s \left(\frac{r}{r_m}\right)^2,
\label{eq:BHth}
\eeqa
where $r>r_m$ and
we consider the adiabatically cooling protons with
the comoving Lorentz factor
$\gamma'_p \sim (\Gamma_m/2\Gamma_s)(r_m/r)$
in Eq.~(\ref{eq:g'p}),
photons with energy
$\varepsilon'_{\rm peak}=(r_b/r)\varepsilon_{\rm peak}$
in Eq.~(\ref{eq:yonetoku}),
and the dissipation radius $r_m/r_b \sim \Gamma_m$ in Eq.~(\ref{eq:rm}).
The ratio in Eq.~(\ref{eq:BHth}) is usually larger than unity.
Therefore, the Bethe-Heitler process is not effective for
a thermal photon spectrum with a cutoff above
the peak energy $\varepsilon_{\rm peak}$.

Whereas, if the photon spectrum is nonthermal,
the protons can interact with high-energy photons to produce pairs.
For a typical Band spectrum,
the protons can cool via Bethe-Heitler processes
even above the pionosphere without $pp$ collisions.
The Lorentz factor of protons decreases to
a value ${\gamma'}^{\rm BH}_p$
for which the Bethe-Heitler cooling optical depth is about unity,
\beqa
n'_{\gamma}\left(\nu'_{\gamma}>2 m_e c^2/{\gamma'}^{\rm BH}_p\right) 
K_{\rm BH} \sigma_{\rm BH} \frac{r_m}{\Gamma_m} \sim 1,
\label{eq:tauBH}
\eeqa
where we estimate at the dissipation radius $r=r_m$ in Eq.~(\ref{eq:rm}), 
and approximate the photon number density above the threshold as
\beqa
n'_{\gamma}\left(\nu'_{\gamma}>2 m_e c^2/{\gamma'}^{\rm BH}_p\right)
=\frac{\left(2 m_e c^2/{\gamma'}^{\rm BH}_p 
\varepsilon'_{\rm peak}\right)^{1-\beta} L}
{4 \pi r_m^2 c \Gamma_m^2 \varepsilon'_{\rm peak} (\beta-1)},
\label{eq:BHngamma}
\eeqa
with the high-energy photon index $\beta \sim 2.5$,
the cross section $\sigma_{\rm BH} \sim
(e^2/\hbar c)(3/8 \pi) \sigma_T$
and the inelasticity $K_{\rm BH}\sim m_e/m_p$ for the
Bethe-Heitler process.
We note that the value $K_{\rm BH} \sigma_{\rm BH}
\sim 5 \times 10^{-31}$ cm$^{2}$ changes
by less than a factor of 3.5 for the range 
$5 \le \nu'_{\gamma} \gamma'_p/m_e c^2 \le 10^3$
\citep{Chodorowski:1992}.
From Eqs.~(\ref{eq:tauBH}) and (\ref{eq:BHngamma}), 
we obtain the Lorentz factor of
protons after the Bethe-Heitler cooling as
\beqa
{\gamma'}^{\rm BH}_p \sim \frac{2 m_e c^2 \Gamma_m}{\varepsilon_{\rm peak}}
\left[
\Gamma_s^{3/2} \eta^{3/2} \eta_{*}^{-4} 
\frac{\varepsilon_{\rm peak}}{m_p c^2} 
\frac{\sigma_{T}(\beta-1)}{K_{\rm BH} \sigma_{\rm BH}}
\right]^{1/(\beta-1)},
\label{eq:g'pBH}
\eeqa
with Eqs.~(\ref{eq:etast}) and (\ref{eq:rm}) and 
$\Gamma_m \varepsilon'_{\rm peak}=\varepsilon_{\rm peak}$,
where ${\gamma'}^{\rm BH}_p$ does not drop to unity in most cases,
differently from the $pp$ collisional case.
We can neglect the Bethe-Heitler cooling above the dissipation radius
as the number density of photons above the threshold
rapidly decreases as 
$n'_{\gamma}(\nu'_{\gamma}>2 m_e c^2/{\gamma'}^{\rm BH}_p)
\propto r^{-(2\beta+1)}$.
The leftover relativistic motions of protons in Eq.~(\ref{eq:g'pBH}) are
converted into the bulk motion via collisionless bulk acceleration.
Therefore, in the case of the effective Bethe-Heitler cooling,
we derive the final coasting Lorentz factor as
\beqa
\frac{\Gamma_c^{\rm BH}}{\eta} 
= \frac{\Gamma_m {\gamma'}^{\rm BH}_p}{\eta}
= \left(\frac{\eta}{\eta_{\rm BH}}\right)^{3/2(\beta-1)},
\label{eq:GcBH}
\eeqa
where
\beqa
\eta_{\rm BH}&=&\Gamma_s^{-(2\beta+1)/3}
\eta_{*}^{8/3}
\left(\frac{\varepsilon_{\rm peak}}{2 m_e c^2}\right)^{2(\beta-1)/3}
\left[\frac{m_p c^2}{\varepsilon_{\rm peak}}
\frac{K_{\rm BH} \sigma_{\rm BH}}{\sigma_T(\beta-1)}\right]^{2/3}
\nonumber\\
&\sim& 6 \times 10^5\ 
\Gamma_s^{-(2\beta+1)/3}
L_{53}^{2/3} r_{b,8}^{-2/3} \varepsilon_{{\rm peak},{\rm MeV}}^{2(\beta-2)/3},
\label{eq:etaBH}
\eeqa
with Eq.~(\ref{eq:etast}).
In Fig.~\ref{fig:etar}, we plot $\Gamma_c^{\rm BH}$ 
of Eqs.~(\ref{eq:GcBH}) and (\ref{eq:etaBH}).
We can see that the Bethe-Heitler process can reduce
the proton kinetic energy by $\sim \Gamma_c^{\rm BH}/\Gamma_c
\sim \Gamma_c^{\rm BH}/\eta
\sim 0.03$--$1$ times (at most)
in the range of the dimensionless entropy
$\eta_{k1} < \eta < \eta_{\rm BH}$.
Nevertheless, the coasting Lorentz factor can attain
a VHLF even if the Bethe-Heitler cooling is most effective.
The actual evolution would be between 
$\Gamma_c$ and $\Gamma_c^{\rm BH}$ in Fig.~\ref{fig:etar},
but is difficult to evaluate exactly
since it depends on the photon spectrum below the photosphere,
which could be thermal or could already be the Band spectrum
(see \S\ref{sec:MeV}).
We also note that the coasting Lorentz factor could be larger than
$\Gamma_c^{\rm BH}$ if the $e^{\pm}$ pairs produced by the 
Bethe-Heitler process make a photosphere above the
coasting radius since the radiative acceleration continues up to
the photospheric radius.

The other photoprocess is the photomeson interaction,
which produces one or more mesons, mostly pions.
The threshold is higher by $m_\pi/m_e \sim 280$,
while the mean cross section 
$K_{pm} \sigma_{pm} \sim 7 \times 10^{-29}$ cm$^{2}$
is $\sim 140$ times higher
than the Bethe-Heitler process
\citep{Chodorowski:1992}.
Then, the ratio of the photomeson cooling to the Bethe-Heitler cooling is
\beqa
\frac{K_{pm} \sigma_{pm}}{K_{\rm BH} \sigma_{\rm BH}} 
\left(\frac{m_\pi}{m_e}\right)^{1-\beta}
\sim 3 \times 10^{-2} \quad {\rm for} \quad \beta=2.5,
\label{eq:pm/BH}
\eeqa
and $\sim 0.5$ for $\beta=2$.
Thus, the photomeson cooling is subdominant
for a typical Band spectrum with $\beta>2$.

The Coulomb collisions with $e^{\pm}$ also
contribute to the proton cooling
because the temperature of $e^{\pm}$ is usually kept at
a nonrelativistic value, much below the proton temperature,
by the Compton cooling
\citep{Beloborodov:2009be}.
The effective radius of the Coulomb cooling is determined by
\beqa
n'_{\pm} \frac{r}{\Gamma} \frac{m_e}{m_p} \sigma_T \ln \Lambda \sim 1,
\eeqa
where $\ln \Lambda \sim 10$ is the Coulomb logarithm.
That is, the Coulomb cooling is effective if
the Thompson optical depth is larger than
$\tau_T > m_p/m_e \ln \Lambda \sim 200$.
Assuming the feasible $e^{\pm}$ density 
in Eq.~(\ref{eq:emin}) at the dissipation radius
$r_m$ in Eq.~(\ref{eq:rm}),
we can estimate the range of the dimensionless entropy $\eta$
where the Coulomb collisions are effective as
\beqa
\eta < \eta_C &\equiv&
\eta_{*}^{16/9} \Gamma_s^{-7/9}
\left[
\frac{4}{3} \frac{\sigma_{pp}}{\sigma_T} \frac{m_e}{m_p} f_{\pm}
\left(\ln \Lambda \right)^2
\right]^{2/9}
\nonumber\\
&\sim&
6 \times 10^4\
L_{53}^{4/9} r_{b,8}^{-4/9} \Gamma_s^{-7/9} f_{\pm}^{2/9},
\label{eq:etaC}
\eeqa
with Eq.~(\ref{eq:g'p})
and the relation
$n'_p=L/4\pi r_m^2 m_p c^3 \eta \Gamma_m$.
Then, for the maximum $e^{\pm}$ multiplicity $f_{\pm}\sim 0.5$
in Eq.~(\ref{eq:fpm}),
we have $\eta_C > \eta_{k1}\sim 2 \times 10^4$ 
in Eq.~(\ref{eq:etak1}) and Fig.~\ref{fig:etar},
that is, the protons cool via Coulomb collisions for $\eta_{k1}<\eta<\eta_C$
even above the pionosphere without $pp$ collisions.
(Note that the $pp$ collisions dominate the Coulomb collisions
for $\eta \siml \eta_{k1}$ even in this case).
The random Lorentz factor of the protons decreases,
reducing the collisionless pressure,
although it may not completely vanish
since the $e^{\pm}$ density also decreases as
the protons, i.e., the energy source of $e^{\pm}$, cool down.
Then, the coasting Lorentz factor $\Gamma_c$ for $\eta_{k1}<\eta<\eta_C$
could not attain the saturation value $\Gamma_c=\eta$
in Eq.~(\ref{eq:Gcall}) and Fig.~\ref{fig:etar}.
Nevertheless, the ratio $\eta_C/\eta_{k1}\sim 3$ is just a factor
and does not depend on the parameters 
so much ($\eta_C/\eta_{k1} \propto L^{2/45} r_b^{-2/45} \Gamma_s^{-1/5}$).
Thus, the critical entropy $\eta_{k1}$ for the $pp$ collisions
in Eq.~(\ref{eq:etak1})
is still a good indicator of the VHLF fireball formation.

In addition to the above two-body processes,
the collective plasma processes could dissipate the 
relativistic proton energy.
At least, a fraction ($\epsilon_B \sim 0.01$) of the proton energy
could be converted into the magnetic field
via the Weibel instability \citep{Medvedev:1999tu}
or the macroscopic turbulence \citep{Goodman:2007ar,Sironi:2007as,Zhang:2008wn}.
However, the conversion into the magnetic field does not reduce
the coasting Lorentz factor
because the magnetic field also provides the relativistic pressure
to expand the fireball (see \S\ref{sec:mag}).
Rather, it opens up a way to the VHLF
since the magnetic pressure can persist even after 
the proton cooling via two-body processes.
On the other hand, the dissipation of the proton energy into electrons
is harmful for making the VHLF.
However, the energy conversion fraction is usually less than half 
($\epsilon_e < 0.5$)
\citep{Spitkovsky:2007zy,Toma:2007hz}.
The plasma dissipation would be also suppressed
after the protons are isotropized.
Therefore, the plasma processes are unlikely
obstacles to the VHLF formation.

\subsection{Consistency with previous works}\label{sec:check}

We have discussed a physical mechanism of 
the collisionless bulk acceleration to create
VHLF fireballs with $\Gamma_c > \eta_{*} \sim 10^3$ 
up to $\Gamma_c \siml 10^6$
in the previous sections.
Since such a VHLF is somewhat extreme and has not been considered seriously,
we examine
whether a VHLF is allowed by the previous observations.

(1) A high-energy cutoff due to $e^{\pm}$ creation provides
information on the bulk Lorentz factor of a fireball
\citep{Baring:1997am,Lithwick:2000kh,Razzaque:2004cx,Murase:2007ya,Aoi:2009ty}.
This method usually gives a lower limit on the bulk Lorentz factor,
which is about $\Gamma_c > 10^2$--$10^3$ so far
\citep{Abdo:2009a,Abdo:2009pg,LAT:2010us}
and consistent with a VHLF
(see \S\ref{sec:090926} for GRB 090926).
Further observations are needed,
and hopefully, more elaborate observations are necessary
because the exponential cutoff is 
usually smoothed to a broken power-law
by multizone effects,
which make it difficult to identify the cutoff
\citep{Aoi:2009ty,Li:2008ub,Granot:2007gn,Bosnjak:2008bd}.

(2) For the internal shocks to take place before the external shock,
the minimum Lorentz factor in the flow has to be below
$\Gamma \siml 3 \times 10^4$.
\citep{Rees:1994nw,Sari:1996we}
This is because the internal shock radius becomes larger
for higher $\Gamma$,
\beqa
r_{\rm sh}=2 \Gamma^2 c t_v
\sim 6 \times 10^{15}\ {\rm cm}\
\Gamma_{4}^2 t_{v,-3}.
\eeqa
On the other hand, the afterglow starts
when the reverse shock crosses the shell.
Since a VHLF evolution is in the so-called thick shell case,
the reverse shock crosses at the duration time $T$.
After that, the hydrodynamic evolution enters into the self-similar phase,
which may be described by the adiabatic condition
$E \sim (4\pi/3) \gamma^2 r^3 n m_p c^2$.
Eliminating $\gamma$ with $T \sim r_{\rm ex}/4 \Gamma^2 c$, 
we obtain the reverse shock crossing radius as
\beqa
r_{\rm ex} \sim 7 \times 10^{16}\ {\rm cm}\ 
E_{53}^{1/4} n^{-1/4} T_{1}^{1/4},
\eeqa
which do not depend on $\Gamma$.
We note that a VHLF with $\Gamma \simg 3 \times 10^4$
is allowed if it is decelerated by internal shocks with slower shells
before external shocks.
This is even favorable for efficient internal shocks
\citep{Kobayashi:1997jk,Kobayashi:2001iq,Beloborodov:2000nn},
and also for the GeV onset delay in \S\ref{sec:delay}

(3) By identifying the peak time of the afterglow light curve
with the decelerating time of the ejecta,
we can constrain the Lorentz factor
\citep{Sari:1999kj,Rykoff:2009ma,Ghirlanda:2009mj,Liang:2009zi}.
Several results do not imply VHLF values but typically $\Gamma \sim 100$--$600$.
However, this method is only applicable to the so-called thin shell case,
whereas a VHLF evolution is likely a thick shell case.
In the thin shell case, the peak time of the afterglow light curve
is expected to be later than the prompt emission,
while in the thick shell case, the peak time is 
comparable to the prompt duration and difficult to observe.
This method also gives a lower limit on the Lorentz factor if
the flow is decelerated by internal shocks with slower shells
before external shocks.

(4) Reverse and forward shock emission in the early afterglows
constrains the Lorentz factor via spectral and temporal modelings
\citep{Sari:1999iz,Zhang:2003wj}.
Several results do not imply the VHLF.
However, the early afterglow modelings are confronted with 
difficulties to interpret the steep and shallow decay 
discovered by {\it Swift} (see also \S\ref{sec:early})
\citep{Zhang:2005fa,Ioka:2005zj,Panaitescu:2006yj,Zhang:2006uj,
Huang:2006ur,Sato:2006jg}.
This method also gives a lower limit on the Lorentz factor if
the flow is decelerated by internal shocks with slower shells
before external shocks.

(5) We can infer the Lorentz factor by
identifying the spectral peak $L(\nu=\varepsilon_{\rm peak})$
with the thermal emission component
\citep{Pe'er:2007xm,Ryde:2009wn}.
Pe'er et al. and Ryde et al. applied this method to several bursts
including GRB 090902B and suggested $\Gamma \sim 300$--$800$ below a VHLF.
However, this method potentially has two solutions,
one of which may provide a low Lorentz factor,
whereas the other could be the real case with a VHLF.
To be more precise, this method initially discriminates two possible cases,
$r_{\rm ph}>r_s$ and $r_{\rm ph}<r_s$,
where it is possible to determine 
the Lorentz factor only in the case of $r_{\rm ph}>r_s$.
This case also provides a consistency inequality 
($\eta<\eta_{*}$) between observables,
so one might think that this is the solution
if the inequality is satisfied.
However, it is logically not a sufficient condition
but just a necessary condition to satisfy the inequality,
so that we cannot exclude the other case $r_{\rm ph}<r_s$,
which allows a VHLF.

(6) Zhang and Pe'er \citep{Zhang:2009aca,Fan:2009gz} claim
that the predicted thermal component 
is not consistent with the observation,
suggesting that the outflow is not radiation-dominated
but Poynting-dominated.
They estimate the maximum temperature allowed by
the observation of GRB 080916C as $T_{{\rm ph},\max}^{\rm ob}=50$ keV,
using the relation $r_b=c \delta t^{\rm ob}$ with
the observed variability time $\delta t^{\rm ob}=0.5$ s,
and conclude that this is below 
and contradicts with the observed peak energy.
However, the central engine size may 
be smaller than $c \delta t^{\rm ob}$, so that
we can raise the maximum temperature 
with $T_{{\rm ph},\max} \propto r^{-1/2}$ to fit the observation.
In addition, the photosphere model usually assumes 
that the black body spectrum is modified to a Band spectrum by
Comptonization via magnetic reconnection,
neutron collisions or repeated shocks,
\citep{Thompson:2006fp,Meszaros:1999gb,Ioka:2007qk,
Beloborodov:2009be,Giannios:2006jb,Giannios:2007yj} 
although the actual mechanism has not yet been revealed
(see also \S\ref{sec:MeV})
\citep{Ioka:2007qk}.
Then, it is possible to reproduce the observed spectrum
by the photospheric emission.

(7) Zou and Piran \citep{Zou:2009ax}
gave an upper limit on the Lorentz factor by requiring that
the observed deep minimum in the prompt phase should be
above the early external shock emission.
The results are $\Gamma \siml 10^3$, not a VHLF.
Since they present an analysis only for the thin shell case,
it is desirable to calculate the thick shell case,
which is the likely case for the VHLF evolution.
This method also gives a lower limit on the Lorentz factor if
the flow is decelerated by internal shocks with slower shells
before external shocks.

In conclusion, the VHLF fireballs are currently consistent with
previous observations.

\subsection{Connection with magnetic acceleration}\label{sec:mag}

A VHLF can also be achieved by the magnetic acceleration of a fireball.
If the energy density is dominated by the magnetic fields,
the magnetic pressure expands the fireball 
up to the equipartition of the comoving energy density between 
the magnetic fields and matter (protons and/or electrons)
\citep{Meszaros:1996ww}.

The difference between the magnetic acceleration and 
the collisionless bulk acceleration is in
the radiation fraction $L_{\rm ph}/L$ after the evolution.
In the collisionless bulk acceleration, 
the radiation fraction $L_{\rm ph}/L$ is about unity
for the VHLF fireballs in Table~\ref{tab:lumi}.
Whereas, in the magnetic acceleration,
the radiation fraction $L_{\rm ph}/L$ depends on the initial condition
(i.e., the radiation and $e^{\pm}$ fraction)
and the magnetic reconnection during the expansion,
thereby not determined from the first principle.

We can also consider the magnetic component
in the collisionless bulk acceleration.
Firstly, the magnetic field could be advected 
from the central engine.
In this case, the final radiation fraction $L_{\rm ph}/L$ 
becomes less than unity
because the magnetic component occupies 
a fraction of the final luminosity as a nonradiative component.
The magnetic component from the central engine could be relevant for the
short GRBs as discussed in \S\ref{sec:short}.
Secondly, the magnetic field could be generated
by the relativistic proton component via plasma processes
at the fireball dissipation (see \S\ref{sec:others}).
This does not reduce the final radiation fraction $L_{\rm ph}/L$.
The magnetic pressure could persist even after
the proton cooling, expanding a way to the VHLF.

\section{Fireball spectrum:
hot photosphere--internal--external shock synchrotron model}\label{sec:spec}

In this section, we apply VHLF fireballs to the GRB emission.
In \S\ref{sec:site}, we first discuss the GRB emission site 
from the kinematical point of view without detailed spectral modelings.
In \S\ref{sec:GeV}, we calculate the internal shock synchrotron spectrum
and find that the VHLF models have advantages on providing
a single rising power-law spectrum over $>$7 energy digits
and also the high internal shock efficiency to supply
sufficient energy to the power-law component that is comparable to
the Band component.
In \S\ref{sec:090926}, we discuss the possible origins of the spectral break
at approximately $1.4$ GeV observed in the extra component of GRB 090926,
which was suggested as the $e^{\pm}$ creation cutoff for $\Gamma \sim 600$.
In the VHLF models, the spectral break could be 
the synchrotron cooling break for $\Gamma \sim 10^4$, 
or the maximum synchrotron cutoff,
particularly limited by the dynamical time, for $\Gamma \sim 10^5$.
In \S\ref{sec:delay}, we suggest that 
the GeV onset delay reflects the timescale 
for the baryon loading rate to change
at the dissipation radius,
i.e., the environmental change around the progenitor star 
controls the GeV onset delay.
We can predict the delay time and also its parameter dependences,
which are found to be consistent with the observations.

\subsection{Emission sites}\label{sec:site}

The GRB emission has at least two, and probably three components:
(1) the usual Band component
(a broken power-law with a peak at $\varepsilon_{\rm peak} \sim 0.1$--$1$ MeV),
(2a) high-energy emission ($>100$ MeV) that lasts longer than
the MeV emission, \citep{Hurley:1994,Gonzalez:2003}
showing a power-law decay,
as revealed by Fermi/LAT 
\citep{Abdo:2009b,Abdo:2009pg,Abdo:2009a,Abdo:2010a,LAT:2010us},
(2b) high-energy variable emission ($>100$ MeV) that looks correlated
with the MeV emission with a short rise and decay time scale \citep{Abdo:2009pg}.
Most of the spectra are well fitted by the Band function 
even up to $\sim 10$ GeV, while 
an additional distinct component at $\simg 10$ MeV
is also found, at least in GRB 090510 and 090902B
out of $\sim 12$ Fermi/LAT bursts.
This additional component is fitted by a single power-law that
slightly rises in $\nu F_{\nu} \propto \nu^{0.1}$--$\nu^{0.4}$
and often extends below $\siml 20$ keV
over $\sim 7$ energy digits.
(However, note that no other experiments have confirmed the
low-energy extension of the power-law component.)

\begin{figure}
\centerline{\includegraphics[scale=.35]{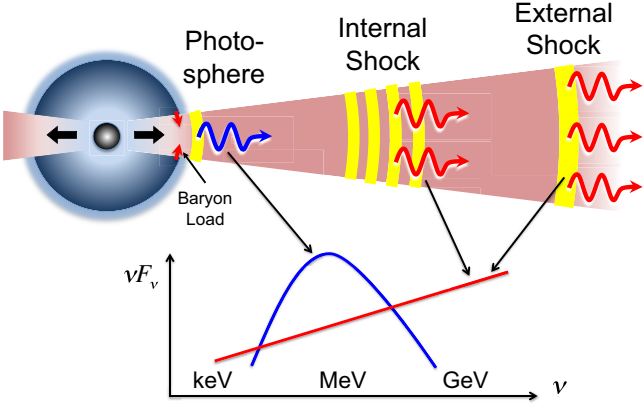}}
\caption{
Photosphere--internal--external shock model \citep{Toma:2010xw}
in which the photospheric emission produces the Band spectrum,
the internal shock contributes the variable power-law spectrum,
and the external shock makes the long-lived power-law spectrum.
The central core of a massive star or
the merged compact star
gravitationally collapses into a black hole or neutron star, 
which somehow launches a collimated jet.
Dissipation via shocks internally or with
nearby matter under the photosphere determines 
the baryon loads and subsequent fireball evolution,
where low baryon loads lead to a VHLF,
as in Fig.~\ref{fig:etar} and Table~\ref{tab:lumi}.
}
\label{fig:model}
\end{figure}

Figure~\ref{fig:model} shows our model in a schematic way.
This is essentially the same as the photosphere--internal shock model
proposed by Toma et al. \citep{Toma:2010xw}
(see \S\ref{sec:GeV} for differences
in the Lorentz factor and hence the emission mechanism).
In this model, we consider the following emission site for each
emission component,
\begin{itemize}
\item[(1)] Band component: Photospheric emission,
\item[(2a)] Long-lived power-law component: External shock,
\item[(2b)] Variable power-law component: Internal shock.
\end{itemize}
These assignments seem reasonable from the kinematical point of view, 
even without detailed spectral modelings,
because of the following reasons:
\begin{itemize}

\item A photospheric origin of the Band component 
easily explains the high radiative efficiency,
in addition to the low-energy spectral index
and the stability of the spectral $\varepsilon_{\rm peak}$ relation
(see \S\ref{sec:intro}).
The radiative efficiency
(i.e., the ratio of the photospheric luminosity $L_{\rm ph}$
to the total luminosity $L$)
can be more than $\sim 50\%$
if the radiation to baryon ratio is in the moderate to high entropy range
$\eta>\eta_{*} \sim 10^3$, 
as summarized in Table~\ref{tab:lumi} with Eq.~(\ref{eq:etast}).
The dissipation under the photosphere is suggested
by the $\varepsilon_{\rm peak}$--$L$ Yonetoku relation
(see \S\ref{sec:base}).

\item An external shock origin of the long-lived power-law component
has an advantage for explaining its longevity and power-law decay.
The high-energy long-lived emission could be consistent with
the simple synchrotron emission from the external forward shock,
with either adiabatic 
\citep{Kumar:2009ps,Kumar:2009vx,Duran:2010et,Corsi:2009ib,Corsi:2009vk,
Gao:2009qa,DePasquale:2009bg,Pandey:2010ti}
or radiative shock 
\citep{Ghirlanda:2009mj,Ghisellini:2009rw}
and probably the Klein-Nishina effect \citep{Wang:2009rp,Nakar:2009er}.
Although the maximum synchrotron frequency has not been detected
\citep{Li:2010zx,Piran:2010ew},
the long-lived component could be produced by the inverse Compton emission
\citep{Zou:2008dr,Wang:2009rp,Corsi:2009ib,Neamus:2010nb,Murase:2009su}.
As we can see from Table~\ref{tab:lumi},
the kinetic energy of the external shock is comparable 
to the total energy if $\eta>\eta_{k1}\sim 10^4 \Gamma_s^{-3/5}$
in the VHLF range and $\eta<\eta_{*}\sim 10^3$
in the low entropy range
with Eqs.~(\ref{eq:etast}) and (\ref{eq:etak1}).

\item An internal shock origin 
is naturally invoked for the variable high-energy component
because the external shock cannot usually produce fast variability
\citep{Sari:1997kn,Ioka:2004gy}.
The external shocks cannot have
the fast decay of the light curve \citep{Ioka:2004gy},
although the fast rise could be produced by
the finite acceleration time 
of particles that radiate at the observed energy
\citep{Duran:2010et}.
It is favorable to have VHLF fireballs in addition to slow ones,
i.e., a large dispersion in the Lorentz factor,
to raise the internal shock efficiency 
for converting the kinetic energy into radiation
\citep{Kobayashi:1997jk,Kobayashi:2001iq,Beloborodov:2000nn},
because the variable power-law component has a comparable luminosity
to the Band component in several bursts.
It is interesting to note a sharp rise of the coasting Lorentz factor
$\Gamma_c$ at $\eta=\eta_{k1} \sim 10^4 \Gamma_s^{-3/5}$ 
in Fig.~\ref{fig:etar}, where
a slight change in the baryon loads leads to
a large difference in the Lorentz factor
$\Delta \Gamma_c/\Gamma_c > 10$.

\end{itemize}

We note that the bulk Compton emission is another possibility
to produce the extra GeV component in the VHLF models.
Since the emission depends on the ambient photon fields,
we leave it as a future work.

\subsection{GeV spectrum: internal shock synchrotron model}
\label{sec:GeV}

A VHLF could be a key for internal shocks to make the
variable power-law spectrum that is slightly rising 
as $\nu F_{\nu} \propto \nu^{0.1}$--$\nu^{0.4}$,
sometimes over $\sim 7$ energy digits
from $\simg 10$ GeV down to $\siml 20$ keV.
In the following, we propose 
the simple internal shock synchrotron model with a VHLF
for the extra spectral component.

The internal shocks convert the kinetic energy into internal energy
with the energy density of
\beqa
U'=\frac{L}{4 \pi r_{\rm sh}^2 c \Gamma^2}
=\frac{L}{16 \pi c^3 t_v^2 \Gamma^6},
\eeqa
where the shock at a radius $r_{\rm sh}=2 c \Gamma^2 t_v$
makes a variability of time $t_v$.
We assume that electrons are accelerated in the internal shock 
to a power-law distribution of Lorentz factor
$\gamma'_e$, 
$d n'_e/d\gamma'_e \propto {\gamma'_e}^{-p}$ 
for $\gamma'_e \ge \gamma'_m$ and $p>2$.
As we have discussed in \S\S\ref{sec:radius} and \ref{sec:others},
abundant $e^{\pm}$ pairs likely exist.
We parametrize the $e^{\pm}$ number density as
\beqa
n'_e={\cal R} n'_p = {\cal R} \frac{U'}{{\bar \gamma}' m_p c^2},
\eeqa
where ${\bar \gamma}'$ is the random Lorentz factor of protons.
If a fraction of internal energy goes into the electron acceleration,
$U'_e= \epsilon_e U'$, the minimum electron Lorentz factor is given by
\beqa
\gamma'_m = \epsilon_e \frac{p-2}{p-1} \frac{m_p}{m_e} 
\frac{{\bar \gamma}'}{\cal R}
\sim 300\ 
\epsilon_e {\bar \gamma}' {\cal R}^{-1} f_p,
\eeqa
where $f_p=6(p-2)/(p-1)$.
We further assume that a fraction of internal energy goes into 
the magnetic field amplification,
\beqa
B'=\left(8 \pi \epsilon_B U'\right)^{1/2}
\sim 4\ {\rm G}\
\Gamma_{4}^{-3} L_{53}^{1/2} \epsilon_{B,-2}^{1/2} t_{v,-3}^{-1}.
\eeqa
Thus, the electron synchrotron cooling is effective above
\beqa
\gamma'_c &=& \frac{6 \pi m_e c}{\sigma_T B'^2 \Gamma t_v}
\sim 4 \times 10^6\ 
\Gamma_{4}^5 L_{53}^{-1} t_{v,-3} \epsilon_{B,-2}^{-1}.
\label{eq:g'c}
\eeqa
With the synchrotron formula 
$\nu(\gamma'_e) = {3 q_e B'} \Gamma {\gamma'_e}^{2}/{4\pi m_e c}$,
we have the synchrotron characteristic and cooling frequencies as
\beqa
\nu_m &=& 50\ {\rm eV}\ 
\Gamma_{4}^{-2} L_{53}^{1/2} t_{v,-3}^{-1} 
\epsilon_e^2 \epsilon_{B,-2}^{1/2} {\bar \gamma}'{}^{2}
{\cal R}^{-2} f_p^2,
\label{eq:num}
\\
\nu_c &=& 9\ {\rm GeV}\ 
\Gamma_{4}^{8} L_{53}^{-3/2} t_{v,-3} \epsilon_{B,-2}^{-3/2},
\label{eq:nuc}
\eeqa
respectively. Because of the slow cooling $\nu_m<\nu_c$ 
for a VHLF, the spectrum is
\beqa
\nu F_{\nu}^{\rm syn} \propto
\left\{
\begin{array}{ll}
\nu^{4/3}, & \quad \nu < \nu_m \\
\nu^{(3-p)/2}, & \quad \nu_m < \nu < \nu_c \\
\nu^{(2-p)/2}, & \quad \nu_c < \nu,
\end{array}\right.
\label{eq:spec}
\eeqa
where the synchrotron luminosity at $\nu=\nu_c$ is about
\beqa
\frac{L_{\rm syn}(\nu_c)}{L}
=\epsilon_e\left(\frac{\gamma_m'}{\gamma_c'}\right)^{p-2}
=\epsilon_e\left(\frac{\nu_m}{\nu_c}\right)^{(p-2)/2},
\label{eq:lumisyn}
\eeqa
since electrons above $\gamma'_c$ cool effectively.

\begin{figure}
\centerline{\includegraphics{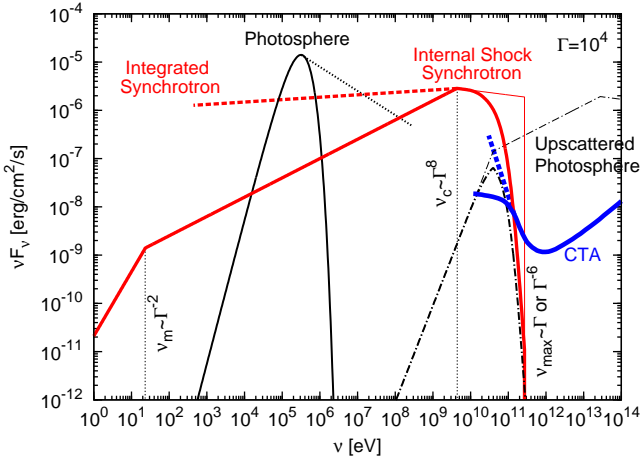}}
\caption{
Spectrum for the internal shock synchrotron model
with a VHLF $\Gamma=10^4$,
for fiducial parameters,
$L=10^{53}$ erg s$^{-1}$,
$\epsilon_e=1$,
$\epsilon_B=10^{-2}$,
$t_v=10^{-3}$ s,
${\cal R}=10$,
${\bar \gamma}'=10$,
$p=2.2$,
and redshift $z=1$,
with ({\it thick solid line})
and without ({\it thin solid line})
the attenuation by 
the cosmic infrared background \citep{Kneiske:2003tx}.
Thanks to the dependences on the Lorentz factor
of the cooling frequency $\nu_c \propto \Gamma^8$
and of the characteristic synchrotron frequency $\nu_m \propto \Gamma^{-2}$,
the rising segment of the $\nu F_{\nu} \propto \nu^{(3-p)/2}$
spectrum is stretched over $> 7$ energy digits.
The maximum synchrotron cutoff is determined by
the cooling time $\nu_{\max}^{\rm cool} \propto \Gamma$ in Eq.~(\ref{eq:numax1})
or by the dynamical time
$\nu_{\max}^{\rm dyn} \propto \Gamma^{-6}$ in Eq.~(\ref{eq:numax2}),
which could produce the $\sim$GeV cutoff 
as observed in GRB 090926 for a VHLF case, $\Gamma \sim 10^{4}$--$10^{5}$.
The spectral integration over continuous regions and/or times
would lead to a flat spectrum in Eq.~(\ref{eq:spec2})
({\it dashed line}).
We also show the photospheric spectrum 
({\it solid line})
with a Comptonized component ({\it dotted line}),
and the upscattered photospheric spectrum
with ({\it thick dot-dashed line}) and without ({\it thin dot-dashed line})
the attenuation.
These are compared with the CTA sensitivity
where we raise the public one by (50 hr/10 s) for simplicity
(as the worst case).
The actual sensitivity in low energy
could be worse than the public one \citep{teshima}.
}
\label{fig:spec}
\end{figure}

Figure~\ref{fig:spec} shows the internal shock synchrotron spectrum
with a VHLF, $\Gamma=10^4$,
for fiducial parameters,
$L=10^{53}$ erg s$^{-1}$,
$\epsilon_e=1$,
$\epsilon_B=10^{-2}$,
$t_v=10^{-3}$ s,
${\cal R}=10$,
${\bar \gamma}'=10$,
$p=2.2$,
and redshift $z=1$.
Here, the attenuation by the $e^{\pm}$ creation with
the cosmic infrared background becomes important at $\simg 20$ GeV,
which is taken into account using 
the best-fit model of Kneiske et al. (2004) \citep{Kneiske:2003tx}.

From Fig.~\ref{fig:spec}, we can find that 
the VHLF internal shock synchrotron model has the following advantages.
\begin{itemize}
\item[(1)] First, the extra component has
a single rising power-law spectrum over many energy digits.
Thanks to the strong dependence of $\nu_c$ on the Lorentz factor,
$\nu_c \propto \Gamma^8$,
a VHLF internal shock has a high cooling frequency $\nu_c$
beyond $\nu_c \simg 10$ GeV.
The $\Gamma$ dependence is $\nu_m \propto \Gamma^{-2}$
for the characteristic frequency $\nu_m$,
extending a power-law below $<1$ keV.
In combination, the rising segment of the $\nu F_{\nu} \propto \nu^{(3-p)/2}$
spectrum is stretched to
\beqa
\frac{\nu_c}{\nu_m}
\sim 2 \times 10^{8} \left(\frac{\Gamma}{10^4}\right)^{10}
\eeqa
for fiducial parameters with Eqs.~(\ref{eq:num}) and (\ref{eq:nuc}).
Therefore, a VHLF could be crucial for making a single power-law spectrum
that is rising over $>$7 energy digits.

\item[(2)] Second, the VHLF models may have the high internal shock efficiency,
which can supply sufficient energy to the extra component that is comparable
to the photospheric Band component.
From Eq.~(\ref{eq:lumisyn}),
the luminosity of the high-energy component
is almost equal to the photospheric one,
$L_{\rm syn}(\nu_c) \sim L_{\rm ph} \sim L$,
if $p\approx 2$ and $\epsilon_e \approx 1$.
The high-electron-energy fraction $\epsilon_e \approx 1$
may be realized by highly efficient internal shocks 
between VHLF and slow shells,
and also by $e^{\pm}$-rich shells
expected for fireballs 
after dissipation (see \S\S\ref{sec:radius} and \ref{sec:others}).
\end{itemize}

Alternatively, the power-law spectrum may be shaped 
by the superposition of emission
from continuous regions and/or times.
Since the dependence of the cooling frequency $\nu_c$ 
on the Lorentz factor $\Gamma$ is strong,
a slight change in $\Gamma$ results in 
a large shift of $\nu_c$,
and hence, in a relatively flat $\nu F_{\nu}$ spectrum.
As an example, we assume
$L \propto \Gamma^{\lambda}$ and $\bar \gamma \propto \Gamma^{1/2}$
for simplicity.
Then, with $\nu_c \propto \Gamma^8 L^{-3/2}$
and $L_{\rm syn}(\nu_c) \propto L (L {\bar \gamma} \Gamma^{-5})^{p-2}$
in Eqs.~(\ref{eq:num}), (\ref{eq:nuc}), and (\ref{eq:lumisyn}), 
the envelope of the spectrum integrated over continuous regions
and/or times becomes relatively flat as
\beqa
\Sigma \left(\nu F_{\nu}^{{\rm syn}} \right)
\propto \nu^{\frac{2\lambda+(\lambda-9)(p-2)}{16-3\lambda}}
\sim \nu^{0.05} \quad {\rm for} \quad \nu < \nu_c,
\label{eq:spec2}
\eeqa
where the last equality is for $p=2.2$, and $\lambda=1$, i.e.,
the case when the luminosity is proportional to the kinetic energy,
which may be reasonable.

Our model predicts a temporal correlation
between the extra power-law component in the low ($\sim$keV)-energy region
and the high ($\sim$GeV)-energy region,
because a single emission mechanism produces the whole power-law component.
In the integrated synchrotron case in Eq.~(\ref{eq:spec2}),
the low-energy component could be delayed
by the variability time, i.e., 
the dynamical time for the Lorentz factor to change.
The photospheric Band component is also expected to be temporally
correlated with the power-law component,
at least within the variability time,
because the photospheric luminosity
is comparable to the kinetic one for the VHLF range,
as shown in Table~\ref{tab:lumi},
and the internal shock emission is delayed by 
the variability time at most.

Toma et al. (2010) \citep{Toma:2010xw} showed that the photospheric
emission can be efficiently scattered by the electrons
in the internal shocks, and the Compton upscattered photospheric
emission can explain
the observed distinct high-energy component.
The low-energy part of the power-law component at $\siml 20$ keV
is attributed to the synchrotron emission, i.e., 
a different origin from the high-energy part.
This is different from the VHLF models that
employ only a single emission mechanism
(synchrotron emission)
and may be favorable for explaining
a single power-law component over $>$7 energy digits.
In the VHLF models,
the upscattered emission is beyond the Fermi energy range,
because the characteristic frequency of upscattered photons is
$\sim {\gamma'}_{m}^2 \varepsilon_{\rm peak} > 1$ TeV
(see Fig.~\ref{fig:spec}),
which is rather a nice target for 
the future Cherenkov Telescope Array (CTA) Project.
The electron cooling by the Compton upscattering
is also suppressed by the Klein-Nishina effect.
The emissivity ratio of the upscattered photosphere to the synchrotron is
\beqa
Y_{\rm up}(\gamma'_e)=\frac{P_{\rm up}(\gamma'_e)}{P_{\rm syn}(\gamma'_e)}
\simeq \frac{U'_{\rm ph}\left[\nu<\nu_{\rm KN}(\gamma'_e)\right]}{U'_B},
\eeqa
where
$\nu_{\rm KN}(\gamma'_e) = {\Gamma_{\rm sh} m_e c^2}/{\gamma'_e}$
and the step function approximation in the second equality
is appropriate
if $d\ln U'_{\rm ph}/d\ln \nu < 2$ for $\nu > \nu_{\rm KN}$
\citep{Wang:2009rp,Nakar:2009er}.
Since the comoving energy density of photospheric photons
is approximated by 
\beqa
U'_{\rm ph}\left[\nu<\nu_{\rm KN}(\gamma'_e)\right] 
\simeq U' \left[\frac{\nu_{\rm KN}(\gamma'_e)}
{\varepsilon_{\rm peak}}\right]^{2-\alpha},
\eeqa
for $\nu_{\rm KN} < \varepsilon_{\rm peak}$,
the upscattering dominates the synchrotron
in the electron cooling $Y_{\rm up}(\gamma'_{Y})>1$
at $\gamma'_e < \gamma'_Y$ where
\beqa
\gamma'_{Y} \sim \frac{\Gamma m_e c^2}
{\varepsilon_{\rm peak}}
\epsilon_B^{-1/(2-\alpha)}
\sim 5 \times 10^{5}\ 
\Gamma_{4} \varepsilon_{{\rm peak},{\rm MeV}}
\epsilon_{B,-2}^{-1/(2-\alpha)}
\eeqa
for $\alpha=1$.
This is lower than $\gamma_c$ in Eq.~(\ref{eq:g'c}),
and hence,
the electrons mainly cool via synchrotron, not 
via the inverse Compton, for the VHLF models.
(Note that we have to take the Compton cooling into account
when we precisely calculate the low-energy end of the 
integrated spectrum in Eq.~(\ref{eq:spec2}),
which is produced by the decelerated shocks.)
We can also neglect the SSC emission 
\citep{Pilla:1997jm,Sari:2000zp,Fan:2007vz,Zou:2008dr,Nakar:2009er,Wang:2009rp},
since the synchrotron photon density is lower and
the peak energy is higher than those of the photospheric emission.

The synchrotron component could also be detected in the optical band. 
It could explain the bright optical prompt emission in some GRBs, 
which is brighter than the extrapolation of the Band component 
\citep{Yost:2007pt,Racusin:2008pd}.

\subsection{GRB 090926: $e^{\pm}$ creation cutoff, cooling break or maximum synchrotron cutoff?}\label{sec:090926}

Recently, a spectral break at approximately $1.4$ GeV has been found 
in the extra power-law component of GRB 090926.
We consider the possible origins of the break in the following.

(1) The $\gamma \gamma$ annihilation with $e^{\pm}$ creation 
leads to a spectral break 
\citep{Baring:1997am,Lithwick:2000kh,Razzaque:2004cx,Murase:2007ya,Aoi:2009ty}
at
\beqa
\nu_{\gamma\gamma}
\sim 20\ {\rm GeV}\ 
\left[
\xi_{-1}^{-1}
L_{53}^{-1}
\Gamma_{3}^{2+2\beta}
t_{v,-3}
\varepsilon_{{\rm peak},{\rm MeV}}^{2-\beta}
(\beta-1)
\right]^{\frac{1}{\beta-1}},
\eeqa
which is determined by the optical depth condition
$\tau(\nu) \sim \xi(\beta) n'_{\gamma} (\nu'_{\gamma}>\tilde \nu') 
\sigma_T r_{\rm sh}/2 \Gamma=1$,
where $\tilde \nu'=m_e^2 c^4/\nu'$,
$\xi(\beta) \simeq 7(\beta-1)/[6 \beta^{5/3}(\beta+1)]
\sim 0.1$, \citep{svensson:1987,Murase:2007ya}
and we approximate the photon number density as
\beqa
n'_{\gamma} (\nu'_{\gamma}>\tilde \nu') = 
\frac{(\tilde \nu'/\varepsilon'_{\rm peak})^{1-\beta} L}
{4\pi r_{\rm sh}^2 c \Gamma^2 \varepsilon'_{\rm peak} (\beta-1)},
\eeqa
with $\beta=2.5$.
Thus, the break at $\sim 1.4$ GeV 
suggests the bulk Lorentz factor of $\sim 600$.
This constraint is applicable to the outermost shell,
not excluding VHLF shells behind the slower shell.
(Note that this alignment is even preferred by the considerations
of the GeV onset delay in \S\ref{sec:delay}).
The $e^{\pm}$ creation cutoff disappears
if $\tilde \nu'_{\gamma \gamma}<\varepsilon_{\rm peak}$, that is,
\beqa
\Gamma > \Gamma_{\gamma \gamma}&=&
\left[\frac{\xi(\beta) L \sigma_T}
{16 \pi c^2 t_{v} \varepsilon_{\rm peak} (\beta-1)} \right]^{1/4}
\nonumber\\
&\sim& 3 \times 10^3\ 
L_{53}^{1/4} t_{v,-3}^{-1/4} \varepsilon_{{\rm peak},{\rm MeV}}^{-1/4},
\eeqa
in the VHLF range,
because the photon density is almost constant below $\varepsilon_{\rm peak}$
for the usual spectral index.
We note that the exponential cutoff is 
usually smoothed to a broken power-law
by multizone effects
\citep{Aoi:2009ty,Li:2008ub,Granot:2007gn,Bosnjak:2008bd}.

(2) The second possibility is the cooling break at
\beqa
\nu_c = 9\ {\rm GeV}\ 
\Gamma_{4}^{8} L_{53}^{-3/2} t_{v,-3} \epsilon_{B,-2}^{-3/2},
\eeqa
in Eq.~(\ref{eq:nuc}).
Thus, the break at $\sim 1.4$ GeV 
suggests a VHLF of $\Gamma \sim 10^4$.
The change in the spectral index at the cooling break
is $0.5$ in Eq.~(\ref{eq:spec}),
which can be used as a test of this possibility,
although the photon number is insufficient in the current observations.
If the low-energy part below the cooling break is produced by
the spectral integration in Eq.~(\ref{eq:spec2}),
the change of the index is smaller than $0.5$.

(3) The third possibility is the maximum synchrotron cutoff,
\beqa
\nu_{\max}^{\rm cool}
=\phi_s \frac{3q_e B'}{4\pi m_e c}
{\gamma'_{\max}}^2 \Gamma
=\phi_{s} \frac{27}{16\pi} \frac{m_e c^3}{e^2}\frac{\Gamma}{\kappa}
=50\ \kappa^{-1} \Gamma_{3}\
{\rm GeV},
\label{eq:numax1}
\eeqa
which is only dependent on the bulk Lorentz factor $\Gamma$,
and is determined by the balance between
the acceleration time
$t'_{\rm acc}=\kappa \gamma'_e m_e c/q_e B'$
and the cooling time $t'_{\rm cool}=3 m_e c/4\sigma_T U'_B \gamma'_e$,
so $\gamma'_{\max}=\left(6 \pi q_e/\kappa \sigma_T B'\right)^{1/2}$,
where the coefficient $\phi_s=0.2294$ is 
quoted from Wijers and Galama (1999) \citep{Wijers:1998st}.
If the break at $\sim 1.4$ GeV is due to the
maximum synchrotron cutoff limited by the cooling time
$\nu_{\max}^{\rm cool}$,
the shock acceleration has to be much slower than the Bohm limit,
$\kappa \sim 50\ \Gamma_{3} \gg 1$, i.e.,
the scattering mean free path is much larger than the Larmor radius.
Apart from the break in GRB 090926, 
the VHLF models ($\Gamma \sim 10^4$) predict the maximum synchrotron cutoff
in the TeV region for $\kappa \sim 1$,
which is a nice target for 
the future Cherenkov Telescope Array (CTA) Project
(see Fig.~\ref{fig:spec}).

On the other hand, if $\nu_{\max}^{\rm cool}<\nu_c$, 
i.e., $\Gamma \simg 10^4$,
the maximum synchrotron cutoff
is limited by the dynamical time $t'_{\rm dyn}=\Gamma t_{v}>t'_{\rm acc}$,
rather than by the cooling time,
yielding $\gamma'_{\max}=q_e B' \Gamma t_v/\kappa m_e c$, and hence,
\beqa
\nu_{\max}^{\rm dyn}=
0.1\ {\rm GeV}\
\Gamma_5^{-6} L_{53}^{3/2} \epsilon_{B,-2}^{3/2} t_{v,-3}^{-3} \kappa^{-2}.
\label{eq:numax2}
\eeqa
Therefore, the break at $\sim 1.4$ GeV could be produced by
VHLF flows with $\Gamma \sim 6 \times 10^4$.
It is a unique feature for the VHLF models to be able to accompany
the maximum synchrotron cutoff limited by the dynamical time,
not by the cooling time.

(4) The last possibility is that
the extra component of GRB 090926
might be the $e^{\pm}$ annihilation line from the photosphere
at the blueshifted energy,
\beqa
\nu_{\pm} = \Gamma m_e c^2
\sim 0.5\ {\rm GeV}\ 
\Gamma_{3},
\eeqa
which is broadened by the order-of-unity distribution 
of the Lorentz factor on the photosphere
\citep{Ioka:2007qk,Murase:2007ya}.
This scenario might be possible
if the $e^{\pm}$ pairs are continuously created on the photosphere,
although the mechanism for the $e^{\pm}$ creation is not apparent.

\subsection{GeV onset delay}\label{sec:delay}

Fermi discovered that the high-energy emission
($>100$ MeV) is delayed behind the onset of the MeV emission
in almost all LAT GRBs.
The delay time in the rest frame is $t_{\rm delay}\sim 1$ s 
for long GRBs and $\sim 0.1$ s for short bursts, 
GRB 081024B and GRB 090510.
These delays are not just caused by the flux increases
above the LAT detection threshold,
but by the spectral changes in the Band and/or extra components
at least in the well-observed bursts.
Since the observed delays of the extra component
are larger than the variability timescale
of the Band component, $\sim 0.01$--$0.1$ s,
the physical origin of the delay
is not likely the kinematic effect \citep{Toma:2010xw}.

In the hot photosphere--internal--external shock model
in \S\ref{sec:site} and Fig.~\ref{fig:model},
the GeV delayed phase arises
when the emission from internal and external shocks 
is weaker than the photospheric emission, i.e.,
almost all the energy escapes in the form of the photospheric luminosity,
not the kinetic luminosity.
This is realized for the dimensionless entropy
in the moderate range $\eta_{*}\sim 10^3 < \eta < \eta_{k1} \sim 10^4$
according to the dissipative hot photosphere model
in Table~\ref{tab:lumi}, Fig.~\ref{fig:etar},
and Eqs.~(\ref{eq:etast}) and (\ref{eq:etak1}).
On the other hand, in the GeV bright phase,
the kinetic luminosity is comparable to the photospheric luminosity,
i.e., in the high-entropy range $\eta_{k1}\sim 10^4 < \eta < \eta_{k3} \sim 10^7$
in Table~\ref{tab:lumi}, Fig.~\ref{fig:etar},
and Eqs.~(\ref{eq:etak1}) and (\ref{eq:etak3}),
which is also the VHLF range.

Therefore, in our picture, the baryon loads decrease progressively
from the GeV delayed phase to the GeV bright phase,
across the critical entropy 
$\eta \sim \eta_{k1}$ in Eq.~(\ref{eq:etak1}) and Fig.~\ref{fig:etar}.
As argued in \S\S\ref{sec:base} and \ref{sec:initial},
the baryon is entrained at the dissipation radius $r_m$,
which is a function of $\eta$ in Eq.~(\ref{eq:rm}).
Then, we can predict the delay time as 
the light crossing time of the dissipation radius
for $\eta=\eta_{k1}$,
\beqa
t_{\rm delay} 
\sim \frac{r_m (\eta_{k1})}{c} 
\sim \frac{r_b}{c} \sqrt{\Gamma_s \eta_{k1}}
\sim 0.5\ s\
L_{53}^{3/5} T_{600{\rm keV}}^{-8/5} \Gamma_s^{1/5},
\label{eq:rmetak1}
\eeqa
where $r_b$ is the base size of the flow in Eq.~(\ref{eq:rb}).
Interestingly, this predicted timescale
is comparable to the observed delay time
$t_{\rm delay} \sim 0.1$--$1$ s.
This coincidence also supports our picture that
the VHLF fireballs are responsible for the extra 
high-energy emission.
In addition, Eq.~(\ref{eq:rmetak1}) combined with the 
$\varepsilon_{\rm peak}$-$L$ Yonetoku relation in Eq.~(\ref{eq:yonetoku})
gives 
\beqa
t_{\rm delay} \sim \frac{r_m(\eta_{k1})}{c} \sim 
\frac{r_{pp}(\eta_{k1})}{c} \sim 
0.5\ s\ L_{53}^{-1/5} \Gamma_s^{1/5},
\label{eq:delay}
\eeqa
which has a weak dependence on the luminosity,
consistent with the observations.
Further observations of the delay time
would discriminate models since 
other models have different parameter dependences.
For example, if the Coulomb collisions control the kinetic luminosity,
the delay time is $t_{\rm delay} \sim r_m(\eta_C) 
\propto L^{-1/6} \Gamma_s^{1/9} f_{\pm}^{1/9}$
with the critical entropy
$\eta_C$ in Eq.~(\ref{eq:etaC}).
The delay caused by the $e^{\pm}$ creation cutoff
also has a different dependence $t_{\rm delay} \propto L \Gamma^{-6}$
\citep{Li:2008ub}.
If we can refine the delay time measurements in the future,
the delay time might be used as a distance indicator
like the $\varepsilon_{\rm peak}$-$L$ Yonetoku relation.

The delay timescale is also comparable to
the light crossing time of the progenitor star
\beqa
t_{\rm delay} \sim \frac{R_{\rm star}}{c} \sim 0.3\ s\
\left(\frac{R_{\rm star}}{10^{10}\ {\rm cm}}\right).
\eeqa
Therefore, a natural picture is that
the fireball dissipation via baryon loads 
is controlled by the environment just outside the star.
As the environment changes with time $\sim R_{\rm star}/c$,
the baryon loads decrease and
the dissipation radius increases beyond the thermalization radius,
leading to a VHLF fireball via the collisionless bulk acceleration
in \S\ref{sec:Gc},
and hence, to the high-energy emission with the onset delay.

Although we have used the light crossing time to estimate the timescales,
the actual environment could be more complex,
which may not be governed by the light speed.
After the jet breakout, the jet is likely surrounded by a cocoon,
which consists of the decelerated jet and the shocked stellar envelope.
The velocity of the cocoon depends on 
the jet luminosity and the stellar structure,
although it is typically $v_c/c \sim 0.1$--$1$
\citep{Toma:2006iu}.
If the accelerated particles are important for the baryon loading,
the typical velocity is the light speed.
The baryon loading process is not clear, and further discussions
are given in \S\S\ref{sec:yonetoku}, \ref{sec:jet}, and \ref{sec:2nd}.

\section{Predictions and open issues}\label{sec:open}

\subsection{TeV neutrino}\label{sec:neutrino}

\begin{figure}
\centerline{\includegraphics[scale=.35]{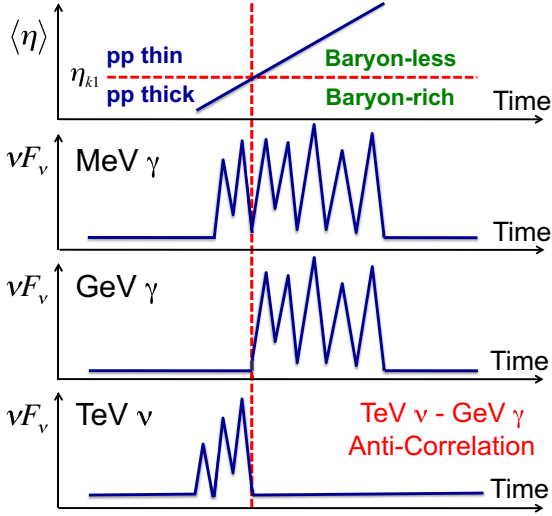}}
\caption{
The anticorrelation between $\sim$TeV neutrinos 
and extra variable GeV $\gamma$-rays
is schematically shown.
This is predicted independently of the neutrino generation processes,
either $pp$ or photomeson interactions,
because the same proton kinetic energy
is shared by neutrinos and extra GeV $\gamma$-rays
in the hot photosphere--internal--external shock model 
in \S\ref{sec:site} and Fig.~\ref{fig:model}.
In our picture of \S\ref{sec:delay}, 
the GeV onset delay phase corresponds to
the baryon-rich (low entropy) phase,
associated with the effective $pp$ neutrino emission
with $\sim$TeV energy in Eq.~(\ref{eq:enu}) 
and the diffuse flux in Eq.~(\ref{eq:nuflux}).
}
\label{fig:neutrino}
\end{figure}

We can predict $\sim$TeV neutrinos and their
temporal anticorrelation
with extra GeV high-energy $\gamma$-rays
in the hot photosphere--internal--external shock model
in \S\ref{sec:site} and Fig.~\ref{fig:model}.
In this model, the baryon (protons) is loaded below the photosphere,
leading to the fireball dissipation 
as suggested by observations
(see \S\ref{sec:base}).
The entrained protons are relativistic in the comoving frame of the
shocked fireball with the Lorentz factor
$\gamma'_p \sim \Gamma_m/2\Gamma_s
\sim \sqrt{\eta/\Gamma_s}/2$ in Eqs.~(\ref{eq:g'p}) and (\ref{eq:rm}).
The $pp$ collisions between these protons produce 
pions, which immediately decay into neutrinos 
\citep{px94,dkk99,Bahcall:2000sa,Meszaros:2000fs}
via
\beqa
\pi^{+} &\to& \mu^{+} + \nu_{\mu} 
\to e^{+} + \nu_e + {\bar \nu}_{\mu} + \nu_{\mu},
\\
\pi^{-} &\to& \mu^{-} + {\bar \nu}_{\mu} 
\to e^{-} + {\bar \nu}_e + \nu_{\mu} + {\bar \nu}_{\mu}.
\eeqa
Each neutrino shares $\sim m_\pi/4 m_p \sim 5\%$ of
the primary proton energy, so that the observed neutrino energy is
\beqa
\varepsilon_{\nu} \sim \frac{m_\pi}{4 m_p}
\Gamma_m \gamma'_p m_p c^2
\sim 0.2\ {\rm TeV}\ \left(\frac{\eta}{10^4}\right).
\label{eq:enu}
\eeqa
The neutrino luminosity originates from the proton kinetic luminosity,
which also produces the extra GeV $\gamma$-ray component in our model.
That is, the same kinetic energy is shared by neutrinos
and extra $\gamma$-rays.
Therefore, the $\sim$TeV neutrinos are predicted to anticorrelate
with GeV $\gamma$-rays, as shown in Fig.~\ref{fig:neutrino}
and Table~\ref{tab:lumi}.
Such a temporal prediction would be interesting
for the upcoming multimessenger astronomy.
The neutrino fluence may be comparable to the MeV $\gamma$-ray fluence
since the kinetic energy is comparable to the photospheric energy,
as shown in Eq.~(\ref{eq:E'm2}) and Table~\ref{tab:lumi}.
However, as discussed in \S\ref{sec:delay},
the $pp$ collision is effective only in the GeV delay phase
with a timescale $\sim t_{\rm delay}\sim 1$ s less than
the total duration $T\sim 20$ s.
Therefore, the neutrino fluence would be $\sim t_{\rm delay}/T \sim 0.05$
times less than the MeV $\gamma$-ray fluence
${\cal F} \sim 10^{-6}$ erg cm$^{-2}$.
With the GRB event rate $R_{\rm GRB} \sim 10^3$ yr$^{-1}$,
the diffuse neutrino background from GRBs is estimated as
\beqa
\varepsilon_{\nu}^2 \Phi_{\nu}
\sim \frac{1}{4 \pi} \frac{t_{\rm delay}}{T} {\cal F} R_{\rm GRB}
\sim 1 \times 10^{-10}\ {\rm GeV}\ {\rm cm}^{-2}\ {\rm s}^{-1}\ {\rm sr}^{-1},
\label{eq:nuflux}
\eeqa
which is less than the current limits 
\citep{Abbasi:2009ig,Abbasi:2009kq,Fukuda:2002}
and less than the IceCube design sensitivity by an order of unity.
If the GeV bright bursts are a minor population 
as the LAT bursts are $\sim 7\%$ of Fermi bursts,
the diffuse neutrino flux is higher by $\sim T/t_{\rm delay} \sim 20$,
i.e., comparable to the diffuse $\gamma$-ray flux from GRBs,
which could be detectable using IceCube in the near future.
We note that IceCube is sensitive to neutrinos above TeV energy,
and not so much to sub-TeV neutrinos in Eq.~(\ref{eq:enu}).
However, a fraction of protons would have the random Lorentz factor
larger than $\gamma'_p$ in Eq.~(\ref{eq:enu}), as discussed 
in \S\ref{sec:initial},
so that the detection is not completely hopeless.

The photomeson interactions ($p\gamma \to n \pi$)
could also generate pions, and hence, neutrinos
if the photon spectrum under the photosphere is already nonthermal,
as discussed in \S\ref{sec:others}.
Each neutrino energy is also 
similar to the $pp$ collisional case
in Eq.~(\ref{eq:enu}),
since the primary proton energies are the same.
We also expect the anticorrelation between
$\sim$TeV neutrinos and GeV $\gamma$-rays in Fig.~\ref{fig:neutrino}
as in the $pp$ collisional case
because the energy source is again the proton kinetic energy, 
which also produces the extra GeV $\gamma$-ray component.
However, the neutrino luminosity is suppressed by
the factor in Eq.~(\ref{eq:pm/BH})
compared with the proton kinetic luminosity
because the Bethe-Heitler process also occurs simultaneously
and consumes the proton energy into $e^{\pm}$ creation.
Therefore, the photomeson neutrinos may be difficult to detect
unless the photon index is hard $\beta<2$ in Eq.~(\ref{eq:pm/BH}).

The neutrino emission could precede the $\gamma$-ray emission
when the jet is still inside the progenitor star 
\citep{Razzaque:2003uv,Pruet:2003yj}.
In this case, the accelerated protons
interact with matter of the progenitor star or synchrotron photons.
The $\gamma$-rays cannot escape owing to the large optical depth.
The time delay between neutrinos and $\gamma$-rays
is expected to be about $R_{\rm star}/v_j \sim 1$--$10$ s,
approximately the time taken
by the jet to emerge from the progenitor star \citep{Toma:2006iu}.

Wang and Dai \citep{Wang:2008zm} and Murase \citep{Murase:2008sp} 
also discussed the high-energy neutrino emission
from the dissipative photospheres of GRBs.
We can predict that 
the high-energy neutrinos in their models would also
temporally anticorrelate with the GeV $\gamma$-rays.

\subsection{MeV $\gamma$-ray spectrum}\label{sec:MeV}

In the photosphere model, it is unclear
how to produce the high-energy nonthermal tail
of the Band spectrum \citep{Ioka:2007qk}.
The main problem is the source of the nonthermal energy
that is comparable to the total energy.
The nonthermal tail has to be produced near the photosphere $\tau_T \sim 1$,
since the spectrum is thermalized above the observed frequency,
\beqa
\nu_{\rm th} \sim \frac{\Gamma m_e c^2}{\tau_T}
\sim 500\ {\rm MeV}\ \left(\frac{\Gamma}{10^4}\right)
\left(\frac{\tau_T}{10}\right)^{-1},
\label{eq:nuth}
\eeqa
through Compton scatterings by the thermal electrons
with temperature $T'_e$ less than that of the nonthermal photons.
This is because a photon changes its energy by
$\Delta \nu'_\gamma/\nu'_\gamma
\sim (4 k T'_e - \nu'_\gamma)/m_e c^2$
in a single scattering,
so that a relatively large optical depth
$\tau_T \sim \nu'_\gamma/\Delta \nu'_\gamma \sim m_e c^2/\nu'_\gamma$
is necessary for thermalization at $\nu'_\gamma> 4 k T'_e$.
Note that the necessary optical depth is not
$\tau_T \sim (\nu'_\gamma/\Delta \nu'_\gamma)^{1/2}$ here,
since the fireball is expanding with decreasing $\tau_T$.

As a by-product of the discussions in the previous sections, 
we could find a hint for the energy source of the high-energy nonthermal
Band spectrum in the photosphere model.
That is the relativistic baryon (protons) component 
loaded below the photosphere
at the fireball dissipation, as suggested by observations
(see \S\ref{sec:base}).
The entrained protons are relativistic in the comoving frame
of the shocked fireball (see \S\ref{sec:initial}), 
and hence, can heat electrons
via $pp$ collisions (see \S\ref{sec:radius}), 
$p\gamma$ (Bethe-Heitler and photomeson) processes,
and Coulomb collisions with 
$e^{\pm}$ created by $pp$ and $p\gamma$ processes
(see \S\ref{sec:others}).
Then, the heated $e^{\pm}$ can produce
the nonthermal broken power-law spectrum
through the unsaturated Comptonization of the thermal photons.
The total proton energy before thermalization is just comparable
to the photospheric radiation energy in Eq.~(\ref{eq:E'm2}),
as required by the nonthermal tail of the Band spectrum.
This energy equipartition comes from the energy and momentum
conservation in Eqs.~(\ref{eq:econs}) and (\ref{eq:pcons}) without fine tuning.
For the electron heating near the photosphere $\tau_T \sim 1$,
the VHLF range $\eta > \eta_{k1}$ may be preferred
because the dissipation radius $r_m$ is near or above
the baryonic photosphere $r_{\rm ph}$ (see Fig.~\ref{fig:etar}).

Alternatively, the plasma turbulence could be initiated
at the baryon loading, at least with mildly relativistic velocities.
Since each fluid element has a relative velocity,
the turbulence would scatter thermal photons
to a broken power-law spectrum,
where the $Y$-parameter can be of order unity
for the dissipation near the photosphere.
The turbulence will be damped within the eddy turnover time.
However, once a nonthermal spectrum is formed, the spectral shape
is almost preserved below $\nu_{\rm th}$ in Eq.~(\ref{eq:nuth}).

Beloborodov (2009) \citep{Beloborodov:2009be}
(see also Ref.~\cite{dkk99})
has recently shown that the $pn$, $pp$, and Coulomb collisions
can heat $e^{\pm}$ to produce the nonthermal Band spectrum
in the context of the neutron-loaded fireballs.
Since the neutron models only produce mildly relativistic nucleons,
it is interesting to calculate the case of the relativistic protons and neutrons
in the shocked fireball frame.
In contrast to the neutron models,
the $p\gamma$ (Bethe-Heitler and photomeson) processes
are also important 
for the relativistic case as shown in \S\ref{sec:others}.
The $p\gamma$ processes have an advantage that
the random Lorentz factor of protons does not drop to unity
in most cases (see \S\ref{sec:others}), and hence, the fireball can realize
both the nonthermal photospheric spectrum
and the fair fraction of kinetic energy
for the internal and external shocks at the same time,
in contrast to the $pp$ collisions.
Kazanas et al. \citep{Kazanas:2002jy,Mastichiadis:2009pd}
also considered the relativistic proton accumulation and
the Bethe-Heitler process
in the supercritical pile model,
which may be relevant to the late evolution above the photosphere.

\subsection{Origin of the
$\varepsilon_{\rm peak}$-$L$ Yonetoku relation}\label{sec:yonetoku}

Thus far, we have just used the $\varepsilon_{\rm peak}$--$L$ Yonetoku relation
in Eq.~(\ref{eq:yonetoku})
as an empirical relation.
Let us consider the physical origin of this relation
in the dissipative hot photosphere model.
As discussed in \S\ref{sec:base}, 
the $\varepsilon_{\rm peak}$--$L$ Yonetoku relation
is reproduced by the fireball dissipation under the photosphere
that is associated with the deceleration of the fireball,
probably via the baryon loading.
The necessary (isotropic) baryon loading rate $\dot M$
can be obtained from Eqs.~(\ref{eq:rb}), (\ref{eq:etadef}), and (\ref{eq:rm}),
without using the Yonetoku relation, as
\beqa
\dot M \sim  \frac{L \Gamma_s}{c^2} \frac{r_b^2}{r_m^2}
\sim 10^{-5}\ M_{\odot}\ {\rm s}^{-1}\
\Gamma_s r_{m,10}^{-2} L_{53}^2 T_{600\rm keV}^{-4}.
\label{eq:mdot}
\eeqa
Interestingly, the dependence $L^2 T^{-4}$
disappears if the $\varepsilon_{\rm peak}$--$L$ Yonetoku relation
in Eq.~(\ref{eq:yonetoku}) is satisfied.
The remaining dependence in Eq.~(\ref{eq:mdot})
is the Lorentz factor of the slow mass $\Gamma_s$ and 
the baryon loading radius (dissipation radius) $r_m$, 
which are all determined by
the environment that supplies the baryon into the jet.
In other words, the $\varepsilon_{\rm peak}$--$L$ Yonetoku relation
is satisfied if the baryon loads are controlled by
the environment and do not differ event by event so much.
This picture is also consistent with the fact that the GeV onset delay is
controlled by the environment in \S\ref{sec:delay}.
However, it is still difficult to derive
the normalization of the baryon loading rate,
partly because the current MHD numerical simulations 
have not yet implemented the key physical processes
of radiation transfer and $e^{\pm}$ creation 
\citep{Lazzati:2009xx,Zhang:2003rp,Mizuta:2004gu,Mizuta:2010gh}.
For Eq.~(\ref{eq:mdot}), the essential relation is
$\Gamma_m \propto \sqrt{\eta}$ in Eqs.~(\ref{eq:Gm2})
and (\ref{eq:rm}), which states that 
the fireballs are relativistic even after the dissipation,
derived from the new Equations~(\ref{eq:econs}) and (\ref{eq:pcons}).

\subsection{Baryon load and jet structure}\label{sec:jet}

As discussed in \S\ref{sec:base}, 
the $\varepsilon_{\rm peak}$--$L$ Yonetoku relation
in Eq.~(\ref{eq:yonetoku})
suggests the fireball dissipation under the photosphere
that accompanies the deceleration of the fireball,
probably via the baryon loading.
However, the actual baryon loading process is unclear,
partly because the current MHD numerical simulations 
have not yet implemented the key physical processes
of radiation transfer and $e^{\pm}$ creation 
\citep{Lazzati:2009xx,Zhang:2003rp,Mizuta:2004gu,Mizuta:2010gh}.

From the kinematical viewpoints,
we may argue that the baryon loading process would require
the nonspherical configuration.
Otherwise, the radiation-dominated outflow from the central engine
with the initial width $r_0 \sim 10^7$ cm
cannot be fully dissipated at the dissipation radius $r_m$
in most cases.
We can show this by comparing $r_m$ with 
the radius where the rapid outflow with width $r_0 \sim 10^7$ cm
completely runs into the shocked region 
with the Lorentz factor $\Gamma_m$ as
\beqa
\frac{r_0 \Gamma_m^2}{r_m}
\sim \frac{\Gamma_m}{r_b/r_0}
\sim \frac{\sqrt{\Gamma_s \eta}}{10\ L_{53}^{1/2} T_{600\rm keV}^{-2}},
\label{eq:causal}
\eeqa
with Eqs.~(\ref{eq:rb}) and (\ref{eq:rm}).
This ratio is larger than unity, i.e.,
the dissipation is not completed at $r_m$,
for almost all parameter regions
$\eta \simg 10^2 \Gamma_s^{-1}$ in Fig.~\ref{fig:etar}.
This problem of the incomplete dissipation
applies to all dissipative photosphere models
with $\Gamma \simg 10^2$ and $L_{\rm ph} \sim L$.
In other words, the one-dimensional picture
of the spherical reverse and forward shocks
is insufficient to explain the GRB observations in the photosphere model.
The picture should be rather close to the complete merger with turbulence.

The nonspherical configuration may be plausible
because the low-entropy fireball is pushed by the high-entropy fireball
and the system is subject to the Rayleigh-Taylor instability
and/or the plasma instabilities
\citep{Waxman:1994hr,Ioka:2007qk}.
Then, the full dissipation condition 
($r_m > \Gamma_m^2 r_0$ in the above) 
would be replaced by
$r_m > \Gamma_s^2 r_0$ that is easily satisfied.
One may also consider a jet configuration of the fireball
that entrains baryon 
(protons and neutrons \citep{Levinson:2003je})
through the boundary of the jet.
If the central engine works intermittently, 
the baryon surrounding the jet, previously
in pressure equilibrium with the jet, 
will enter the funnel sideways.
As long as the jet opening angle $\theta_j (> \Gamma_r^{-1})$
is smaller than the causal angle $\Gamma_m^{-1}$,
the fireball can be fully dissipated
via turbulence, e.g., caused by the Kelvin-Helmholtz instability
\citep{Goodman:2007ar,Sironi:2007as,Zhang:2008wn}.
Note that the jet opening angle remains constant
after the jet reacceleration
because the causal angle shrinks all the time.

Although the causal angle 
$\Gamma_m^{-1} \sim 0.01 \eta_4^{-1/2} \Gamma_s^{-1/2}$
for the high entropy flow with $\eta>10^4$
may be smaller than the conventional opening angle
$\theta_j \sim 0.1$,
the jet could have two components or a continuous structure
with the high entropy flow surrounded by
the low to moderate entropy flow of $\eta \sim 10^2$--$10^4$.
We note that the two-component jet is suggested for
GRB 080319B \citep{Racusin:2008pd}.
If this is the case, 
the extra GeV component associated with the high entropy flow
(see Table~\ref{tab:lumi})
may be observed only from the viewing angle near the jet center,
possibly consistent with the fact that 
not all observed bursts are bright with GeV $\gamma$-rays.

The initial variability time $\sim r_0/c \sim 3 \times 10^{-4}$ s
reflecting the central engine size
could be preserved at the dissipation
since the successive shells may not contact with each other
for $r_m < r_0 \Gamma_m^2$ in Eq.~(\ref{eq:causal}).
If the dissipation is nonspherical as discussed above,
the timescale could instead be determined by
the crossing time of the causal region,
$r_m/c \Gamma_m = r_b/c \sim 3 \times 10^{-3}$ s,
with Eqs.~(\ref{eq:rb}) and (\ref{eq:rm}).

\subsection{Multiple baryon loads}\label{sec:2nd}

The baryon-loaded shell could expand and collide before coasting
against another slow mass
$M_{s2}$ with a Lorentz factor $\Gamma_{s2}$.
This second merger is characterized by the 
dimensionless entropy $\eta_2=E_r' \Gamma_r/M_{s2} c^2$,
since the rapid shell is radiation-dominated as in the first merger.
The merger is divided into three types:
\begin{itemize}
\item[(1)] If $\eta_2 > \Gamma_r^2/\Gamma_{s2}$,
we may neglect the second merger
since the rapid shell is not sufficiently decelerated
and the fireball temperature is almost constant.

\item[(2)] If $\eta_2 < \eta$, the total amount of baryon is determined by the
second merger.
The Lorentz factor of the fireball goes down to
$\Gamma_{m2} \sim \sqrt{\Gamma_{s2} \eta_2}$
and then up to $\Gamma_{c2} \sim \eta_2$
with a different base radius $r_{b2} \sim r_{m2}/\Gamma_{m2}$.
We may consider that the initial condition of the dissipated fireball 
is reset by the second merger.

\item[(3)] If $\eta < \eta_2 < \Gamma_r^2/\Gamma_{s2}$,
the total amount of baryon is not changed so much,
while the fireball is decelerated to have a different temperature
and base radius.
The initial radius of the dissipated fireball is 
not $r_m/r_b \sim \Gamma_m \sim \sqrt{\Gamma_s \eta}$,
but $r_{m2}/r_b \sim \Gamma_{m2} \sim \sqrt{\Gamma_{s2} \eta_2}$
that tends to bring the dotted line upward in Fig.~\ref{fig:etar}.
As a result, we have different critical entropies
from $\eta_{k1}, \eta_{k2}, \eta_{k3}$
in Eqs.~(\ref{eq:etak1}), (\ref{eq:etak2}), (\ref{eq:etak3}),
and therefore, a different maximum Lorentz factor 
from $\Gamma_{c,\max}$, $\Gamma_{c,\max}^{c}$ 
in Eqs.~(\ref{eq:Gcmax1}), (\ref{eq:Gcmax2}).
It is straightforward to obtain these quantities in this case.
\end{itemize}
We note that the last value of the base radius 
determines the observed relation in Eq.~(\ref{eq:rb}).

If the merger type is (3), the necessary baryon loading rate 
becomes different from that in Eq.~(\ref{eq:mdot}).
Therefore, the merger type (3) could introduce an outlier
in the Yonetoku relation,
if the origin of the Yonetoku relation is correct in \S\ref{sec:yonetoku}.
To have a tight Yonetoku relation,
the baryon loading needs to be larger at the outermost radius,
which may be reasonable since
the jet boundary also becomes larger.

\subsection{Early X-ray afterglow}\label{sec:early}

{\it Swift} discovered the steep and shallow decay phase
in the early X-ray afterglows
\citep{Zhang:2005fa,Ioka:2005zj,Panaitescu:2006yj,Zhang:2006uj,
Huang:2006ur,Sato:2006jg}.
The important point of these observations is
that the early X-ray afterglows are too dim 
for the conventional internal--external shock model,
in which the kinetic energy left for the afterglow emission
is comparable to or usually larger than 
the prompt energy released at the internal shocks,
so-called the internal shock efficiency problem \citep{Ioka:2005zj}.
In the hot photosphere--internal--external shock model 
in Fig.~\ref{fig:model} and Table~\ref{tab:lumi},
this is not a problem because
the kinetic energy fraction of the afterglow
may be relatively small for the moderate entropy range 
$\eta_{*} (\sim 10^3) < \eta < \eta_{k1} (\sim 10^4)$
in Table~\ref{tab:lumi},
where the fireball becomes optically thin 
in the accelerating phase
before converting all the radiation energy into kinetic energy.

On the other hand,
the LAT bursts with bright GeV $\gamma$-rays 
do not seem to be associated with 
the steep and shallow X-ray afterglows,
although more events are necessary to confirm this fact.
In these bursts, the kinetic energy fraction of the afterglow 
may be larger than that of the {\it Swift} bursts.
This may be consistent with our picture
that the extra GeV emission is produced by 
the relatively high entropy fireball with $\eta > \eta_{k1} (\sim 10^4)$
in the VHLF range,
whose kinetic luminosity is
comparable to the photospheric luminosity,
as shown in Table~\ref{tab:lumi}.
If the jet is structured with the high entropy flow surrounded by
the moderate entropy flow as discussed in \S\ref{sec:jet},
the shallow decay in {\it Swift} events may be detected for observers
in the direction of the moderate entropy flow,
since the high-entropy region with large kinetic energy is 
progressively seen as the afterglow is decelerated
\citep{Toma:2005uv}.

\subsection{Short GRB: magnetized jet from white dwarf?}\label{sec:short}

Short LAT GRBs 081024B and 090510 have smaller
GeV delay time,
$t_{\rm delay} \sim 0.1$ s, than long GRBs.
According to \S\ref{sec:delay}, 
this implies that the size of the progenitor 
for short GRBs is smaller than that for long GRBs.
Interestingly,
the inferred size $c t_{\rm delay} \sim 10^9$ cm
is comparable to the white dwarf radius.
The white dwarf model may also be favorable to explain
the extended emission observed in the short GRBs,
since the accretion time of the white dwarf material 
into the central engine can be much longer than the neutron star case,
where the neutron star accretion
is too fast to explain the long timescale ($> 100$ s) of the
extended emission
\citep{Barthelmy:2005bx,Norris:2006rw}.

For short GRBs, the LAT ($\sim$GeV) fluence 
is comparable to and even larger than
the GBM ($\sim$MeV) fluence,
in contrast to long GRBs with the LAT-to-GBM fluence ratio of $\sim 0.1$
\citep{Ghisellini:2009rw}.
This may require a modification to the simple version of our model,
because the extra GeV luminosity
is usually less than the photospheric luminosity
in Eq.~(\ref{eq:lumisyn}).
One possibility is that the short GRBs could have a low entropy phase
with $\eta < \eta_{*}$ (baryon-rich phase), in which
the photospheric emission is suppressed below the kinetic one
(see Table~\ref{tab:lumi}).
The other possibility is a magnetized fireball jet because
magnetic fields are not radiated away at the photosphere,
reducing the photospheric emission (see \S\ref{sec:mag}).
The magnetic fields also increase the kinetic energy 
by pushing the matter even above the photosphere
via the magnetic pressure,
which could enhance the GeV emission.


\section{Summary}

We have investigated the fireball model with dissipation
under the photosphere, as suggested by the observed
spectral $\varepsilon_{\rm peak}$--$L$ Yonetoku relation 
in Eq.~(\ref{eq:yonetoku}) (\S\ref{sec:base}).
We find that
the fireball can entrain the relativistic baryon component
at the dissipation, with energy comparable to the radiation component,
as derived from the energy and momentum conservation
in Eqs.~(\ref{eq:econs}) and (\ref{eq:pcons}).
The relativistic baryon (proton) component can alter
the GRB fireball dynamics and spectra in novel ways,
which we have studied in this paper.
Our main results may be summarized as follows.

\begin{itemize}

\item As outlined in \S\ref{sec:idea},
the relativistic baryon component can reexpand
to a very high Lorentz factor (VHLF) $\Gamma \sim 10^3$--$10^6$,
much larger than the conventional upper limit $\Gamma \siml 10^3$,
(\S\ref{sec:Gc}, Figs.~\ref{fig:schematic} and \ref{fig:etar}).
Since the pressure is provided by
the relativistic collisionless motions of protons
(and the magnetic field generated by these protons),
this mechanism may be called the collisionless bulk acceleration.
The VHLF is achieved when the baryon load is low (i.e., high entropy)
without efficient thermalization of protons via
$pp$, $p\gamma$ (Bethe-Heitler and photomeson),
Coulomb, and plasma interactions (\S\S\ref{sec:radius} and \ref{sec:others}).
The kinetic energy can become comparable to
the total energy in the VHLF fireballs
(\S\ref{sec:initial}, Table~\ref{tab:lumi}, Fig.~\ref{fig:etar}),
which can alter the internal and external shock emission (see below).
These VHLF models are currently consistent
with previous observations (\S\ref{sec:check}).

\item The VHLF fireballs can explain the extra variable GeV component
in the GRB spectrum using the simple internal shock synchrotron emission.
In the VHLF models,
a single emission mechanism (synchrotron emission) can make
a rising power-law spectrum over $>$7 energy digits,
and the internal shock efficiency can also be sufficiently high to
produce the emission energy comparable to the total one,
as observed by Fermi/LAT.
The VHLF emission is also a nice target 
for the future Cherenkov Telescope Array (CTA),
since the $e^{\pm}$ creation cutoff goes beyond
the TeV range in the VHLF models.
The main Band component may be attributed to the photospheric emission,
and the long-lived GeV component to the external shock
in this hot photosphere--internal--external shock synchrotron model
(\S\S\ref{sec:site} and \ref{sec:GeV}, Fig.~\ref{fig:model}).

\item In the VHLF models, the spectral break
at $\sim 1.4$ GeV observed in the extra component of GRB 090926
is not caused by the $e^{\pm}$ creation cutoff with $\Gamma \sim 600$
(\S\ref{sec:090926}).
The spectral break could be 
the synchrotron cooling break for $\Gamma \sim 10^4$, 
or the maximum synchrotron cutoff,
particularly limited by the dynamical time
(not by the cooling time) for $\Gamma \sim 10^5$.

\item The observed GeV onset delay can be naturally
explained in the dissipative hot photosphere model
because the baryon loads at the dissipation would be rich
shortly after the jet breakout,
and hence, the relativistic baryon component is thermalized into radiation
in the fireball,
leaving little kinetic energy for the GeV emission
(\S\ref{sec:delay}).
The timescale of the GeV onset delay can be calculated
from the theory as the timescale at the dissipation radius,
i.e., the timescale to change 
the baryon loading rate, and hence,
the environment around the progenitor star.
The predicted delay time $\sim 0.5$ s and also its weak dependence 
on the luminosity $t_{\rm delay} \propto L^{-1/5}\sim L^{-1/6}$
in Eq.~(\ref{eq:delay})
are consistent with the observations.

\item The dissipative hot photosphere model predicts
$\sim$TeV neutrinos in Eq.~(\ref{eq:enu})
and the anticorrelation between $\sim$TeV neutrinos and 
the extra variable GeV $\gamma$-rays, independently of 
the neutrino generation processes, either $pp$ or photomeson
interactions,
because the same proton kinetic energy
is shared by neutrinos and extra GeV $\gamma$-rays.
In the optimistic case,
the diffuse neutrino background from GRBs 
in Eq.~(\ref{eq:nuflux}) could be detected
using IceCube in the near future.

\item The relativistic baryon component loaded into the fireball
at the dissipation could work as the as-yet-unknown energy source 
to deform the photospheric thermal spectrum into 
the nonthermal Band spectrum (\S\ref{sec:MeV}).
Without fine tuning,
the relativistic baryon component has the right amount of energy
(comparable to the thermal energy)
to make the nonthermal tail (\S\ref{sec:initial}).
The protons could heat $e^{\pm}$ via $pp$, $p\gamma$ 
(Bethe-Heitler and photomeson), and Coulomb interactions,
leading to the Comptonization of
the thermal photons into the broken power-law spectrum.
Alternatively, the mildly relativistic
plasma turbulence induced by the baryon loading
could make the nonthermal spectrum.
Further investigations in this direction appear interesting.

\item The spectral $\varepsilon_{\rm peak}$--$L$ 
Yonetoku relation can be reproduced by the fireball dissipation
if the baryon loading rate is determined by the environment
that is nearly identical to any bursts in Eq.~(\ref{eq:mdot}).
Nearly identical environments could be plausible
if GRBs are produced by a certain class of progenitor.
For the derivation of the $\varepsilon_{\rm peak}$--$L$ Yonetoku relation,
it is important that
the fireballs are relativistic even after the dissipation,
as deduced from the energy and momentum conservation
in Eqs.~(\ref{eq:econs}) and (\ref{eq:pcons}).

\item The actual baryon loading process at the dissipation 
has not been clarified.
From the kinematics, the dissipation process would require
the nonspherical configuration,
such as the turbulence and/or the jet boundary (\S\ref{sec:jet}).
In the jet case, the causality argument suggests
a two-component or structured jet.

\item A structured jet with both the radiation-dominated and VHLF flows
could explain the steep/shallow decay of early X-ray afterglows 
observed in the {\it Swift} bursts
(as viewed from the radiation-dominated flow)
and the possible paucity of the steep/shallow decay phase 
in the Fermi/LAT bursts
(as viewed from the VHLF flow) in a unified fashion (\S\ref{sec:early}).

\item We have speculated that
the short GRBs might accompany
the magnetized or baryon-rich jet (for the high GeV fluence ratio)
from the white dwarf progenitor (for the short GeV onset delay)
(\S\ref{sec:short}).

\end{itemize}

\acknowledgments
We thank K.~Asano, N.~Kawanaka, D.~Kazanas,
P.~M\'esz\'aros, A.~Mizuta, K.~Murase,
T.~Nakamura, Y.~Ohira, F.~Takahara, M.~Teshima, K.~Toma, X.~Y.~Wang,
and R.~Yamazaki
for useful discussions.
This work is supported in part
by Grants-in-Aid from the 
Ministry of Education, Culture, Sports, Science and Technology
(MEXT) of Japan, Nos.19047004, 21684014, and 22244019.

%

\end{document}